\documentclass{capsnew}
\usepackage{psfig,latexsym,longtable,lscape,graphicx}

\newcommand{\gtrsim}{\mathrel{\hbox{\rlap{\lower.55ex \hbox {$\sim$}}
                   \kern-.3em \raise.4ex \hbox{$>$}}}}
\newcommand{\lesssim}{\mathrel{\hbox{\rlap{\lower.55ex \hbox {$\sim$}}
                   \kern-.3em \raise.4ex \hbox{$<$}}}}

\DeclareRobustCommand{\ion}[2]{%
\relax\ifmmode
\ifx\testbx\f@series
{\mathbf{#1\,\mathsc{#2}}}\else
{\mathrm{#1\,\mathsc{#2}}}\fi
\else\textup{#1\,{\mdseries\textsc{#2}}}%
\fi}

\begin{document}

\pagenumbering{roman}
\cleardoublepage

\pagenumbering{arabic}

\setcounter{chapter}{9}

\setcounter{table}{0}

\author[Erik Kuulkers, Andrew Norton, Axel Schwope, \&\ Brian Warner]{Erik Kuulkers\\
ESA-ESTEC, SCI-SDG, Keplerlaan 1, 2201 AZ Noordwijk, The Netherlands\\
\&\ Aurora Technology BV, Hoofdstraat 305, 2171 BG, Sassenheim, The Netherlands
\and
Andrew Norton\\Department of Physics \&\ Astronomy, The Open University, Walton Hall, Milton \\Keynes MK7 6AA, United Kingdom
\and
Axel Schwope\\Astrophysikalisches Institut Potsdam, An der Sternwarte 16, 14482 Potsdam, Germany
\and
Brian Warner\\Department of Astronomy, University of Cape Town, Rondebosch 7700, South Africa
}

\chapter{X-rays from Cataclysmic Variables}

\section{Introduction}
\label{intro}

Cataclysmic Variables\footnote{See Hack \&\ La Dous (1993) and Warner (1995) for comprehensive overviews.} 
(CVs) are a distinct
class of interacting binaries, transferring mass from a donor star to a
degenerate accretor, a white dwarf (WD). In all observational determinations, and as is
required by theory for stable mass transfer, the donor star is of lower
mass than the accretor.

The majority of CVs have orbital periods, $P_{\rm orb}$, between 75\,min and 8\,h
(see Ritter \&\ Kolb 2003) 
and consist of Roche Lobe-filling main sequence donors and WDs. 
These are WD analogues of the low-mass X-ray binaries (LMXBs; see Chapter 1). 
In the period range 8\,h--3\,d 
the donors must have larger radii than dwarfs in order to fill their Roche Lobes 
and are therefore
evolved subgiants. A few CVs are found with $P_{\rm orb}$$\sim$200\,d, which require
giant donors for them to be Lobe-filling. The absence of evolved CVs
with periods $\sim$3 to $\sim$200\,d is connected with the dynamical instability
that results from an initial donor that had a mass larger than about
67\%\ of that of the WD; such binaries will have experienced rapid
mass transfer and shortened their periods during a common envelope phase.
Beyond $P_{\rm orb}$$\sim$200\,d, mass-transferring systems also exist. These constitute
the {\it symbiotic binaries} (SBs) and are in general not Roche Lobe-filling, but instead
consist of a WD orbiting in the wind of a supergiant, and are
thus analogues of the high-mass X-ray binaries (see Chapter 1).

At the short end of the period range a different kind of CV exists, i.e., in
which the mass losing donors are themselves WDs; they are entirely
deficient in hydrogen. These helium-transferring CVs are known, after the
type star, as {\it AM\,CVn stars} and are observed to have $P_{\rm orb}$ from 60\,min down
to at least as short as 10\,min. They have evolved by passage through two
common envelope phases, which leaves the cores of both of the component
stars exposed.

The CVs are divided into subtypes. These were originally based entirely on
the behaviour seen in long-term optical light curves, but to this are now
added more subtle parameters such as the presence of polarisation. In
essence, a CV's gross behaviour is determined by the rate of mass
transfer from the donor, $\dot{M}_{\rm donor}$, and the strength of the magnetic
moment, $\mu$, of the WD. Other parameters, such as $P_{\rm orb}$, 
mass ratio, and chemical abundances, have less effect; but the brightness and spectral
variations on an orbit time scale can depend strongly on the inclination.

For $\mu$$\lesssim$10$^{31}$\,G\,cm$^{-3}$ 
(corresponding to a magnetic field strength $B$$\lesssim$10$^4$\,G) 
the WD is essentially ``non magnetic" (but see Sect.~\ref{dno}) and mass is
transferred from donor to WD via an accretion stream and
accretion disk without any significant magnetic influence on the fluid
flow. Even for $\mu$ up to $\sim$10$^{33.5}$\,G\,cm$^{-3}$ a stream and accretion disk can
form, but the inner regions of the disk are removed by the magnetosphere
of the WD, within which fluid flow is magnetically channelled. For
larger $\mu$ no disk can form at all; the stream flow from the donor
couples onto field lines from the WD before the stream can circle
around the WD. These configurations, which apply for typical 
$\dot{M}_{\rm donor}$$\sim$10$^{-10}$--10$^{-8}$\,M$_{\odot}$\,yr$^{-1}$, 
are known as {\it intermediate polars} (IPs) for the systems with
intermediate field strength\footnote{The name `intermediate polar' was introduced 
by Warner (1983).}, and {\it polars} (or {\it AM\,Her stars}, after the type
star) for those with strong fields. The field in polars is so strong that
it couples to the field of the donor and forces the WD to
corotate with the binary; it also prevents 
the formation of an accretion disk. In IPs the WD does not
corotate. The term {\it DQ\,Her star} is often used for IPs of
rapid rotation, typically $\lesssim$250\,s, but some use the term in place of
IP, treating DQ\,Her as the type star.

Where disks exist there are two principal behaviours, resulting from the
different viscosities (and hence ability to transport angular momentum) in
cool disks and hot disks (see Chapter 13). 
For low $\dot{M}$ through the disk, the disk is
lower in temperature and the viscosity is too low to transport mass
through the disk as fast as it arrives from the donor;
this state is referred to as {\it quiescence}.
The quiescent disk (which is not stationary) 
therefore acts as a reservoir of gas, and when a critical density is
reached it becomes optically thick, heats up and increases in viscosity,
and rapidly transfers gas onto the WD. The resultant release of the
gravitational energy of the stored gas gives rise to {\it dwarf nova} (DN) outbursts.
The DN class is subdivided into {\it U\,Gem stars} which have 
outbursts on time scales of weeks, and {\it Z\,Cam stars} where outbursts are
occasionally interrupted by a standstill at a brightness intermediate
between outburst and quiescence.
In systems with short $P_{\rm orb}$ an additional phenomenon occurs --
{\it superoutbursts}, which are typically brighter and five times as long as
normal outbursts, during which humps in the light curves 
(so-called `superhumps') are present with a period of a few \%\ longer than $P_{\rm orb}$.
Superoutbursts are thought to be due to tidal stresses between the
outer disk and donor, adding to the higher viscous stress in the outer disk;
it may even result in additional mass transfer from the donor.
This subtype of DN is known as {\it SU\,UMa stars}.

Outbursts, usually of lower amplitude and short duration, can arise in the
truncated disks of IPs. On the other hand, if $\dot{M}_{\rm donor}$ is
high enough it can maintain the disk at high viscosity, producing an
equilibrium state that does not undergo normal DN outbursts
(though modified outbursts are sometimes seen). These systems are known,
from the appearance of their spectra, as {\it nova-like} systems (NLs).
Those showing absorption lines are called {\it UX\,UMa stars}, and those showing
emission lines are {\it RW\,Tri stars}.

The strongly magnetic systems have no disks and therefore lack outbursts. But it is
common for polars to show states of low luminosity caused by lowering of
$\dot{M}_{\rm donor}$ by as much as two or three orders of magnitude. Other types also
show (occasional) low states, especially NLs with 3$<$$P_{\rm orb}$(h)$<$4,
which are known as {\it VY\,Scl stars}.

It is supposed that all CVs, with the exception of the helium-transferring
AM\,CVn stars, undergo thermonuclear runaways at the base of the accreted
hydrogen-rich layers on the WDs as soon as the layers are massive
enough (typically $\sim$10$^{-4}$\,M$_{\odot}$). These produce {\it classical nova} (CN) 
eruptions and are in some ways the equivalent of X-ray bursts in LMXBs (see Chapter 3).
Systems which have shown nova eruptions more than once are referred to as {\it recurrent novae} 
(RNe), whereas SBs with nova eruptions are known as 
{\it symbiotic novae} (SBNe).

The X-ray behaviours are correlated with the above described CV subtypes, 
but not always in a positive fashion. During a DN outburst
hard X-rays\footnote{With `soft' and `hard' X-rays we generally refer to 
X-rays with energies of order 10\,eV and of order keV,
respectively.}
have been seen to increase at the beginning (often with a delay of up
to a day after the optical outburst begins), but then are suppressed until
near the end of the outburst. Similarly, the high $\dot{M}$ (and, therefore, high
accretion luminosity) NLs have relatively low X-ray luminosities, $L_{\rm X}$.
On the other hand, soft X-ray$^{\ast}$ fluxes are greatly enhanced during DN 
outbursts, again with a possible delay with respect to the optical,
which is caused by the time taken for an outside-in outburst to travel
from the cooler outer parts of the disk to the inner disk and the
WD/disk boundary layer (BL).

The polars have high soft $L_{\rm X}$ and the IPs
have relatively high hard $L_{\rm X}$. All of these different X-ray
behaviours are simply connected with the optical depth of the 
BL in the non-magnetic CVs and with the nature of the channelled
accretion flow in the magnetic CVs. More details are given in subsequent
Sections.

X-rays may also be generated in other parts of CV structures -- e.g., in the
shock waves where gas ejected by nova eruptions meets the interstellar
medium; on the hot WD surface after a nova eruption; and in minor
contributions from the magnetically active regions on the surface of the
donor.

SS\,Cyg, one of the optically brightest DN (m$_{\rm V}$$\sim$12--8), 
was the first CV to be detected in X-rays during 
a rocket flight when it was in outburst (Rappaport et al.\ 1974). 
The spectrum was soft with a black-body temperature, $kT_{\rm bb}$$<$130\,eV.
In its quiescent state it was first detected by {\it ANS}, both in
soft and hard X-rays.
In the soft band, its flux was only a few percent
of that observed during outburst (Heise et al.\ 1978). 
Since then many more CVs have been detected in X-rays.

We here review X-ray observations of CVs, with some emphasis on what has been
achieved in the last decade, up to 2003. 
For earlier, more general, reviews we refer the interested reader to, e.g.,
C\'ordova \&\ Mason (1983), Hack \&\ La Dous (1993), C\'ordova (1995) 
and Warner (1995). 
We note that a lot of CVs have recently been found in globular clusters;
we refer to Chapter 8 for an overview of this subject.

\section{X-ray emission from non-magnetic CVs}
\label{nonmagnetic}

\subsection{General properties}
\label{overview}

The collective X-ray properties of CVs have been the subject of several studies
using observations from {\it Einstein} 
(Becker 1981; 
C\'ordova \&\ Mason 1983, 1984; 
Patterson \& Raymond 1985a; 
Eracleous et al.\ 1991a,b), 
{\it ROSAT} (Vrtilek et al.\ 1994; 
van Teeseling \&\ Verbunt 1994; 
Richman 1996; 
van Teeseling et al.\ 1996; 
Verbunt et al.\ 1997), 
and {\it EXOSAT}/ME (Mukai \&\ Shiokawa 1993). 
With a few exceptions
all non-magnetic CVs\footnote{See, e.g., Verbunt (1996) and Mukai (2000)
for earlier reviews of X-rays from non-magnetic CVs.} 
radiate at X-ray flux levels
$\lesssim$10$^{-11}$\,erg\,cm$^{-2}$\,s$^{-1}$ 
(e.g., Patterson \&\ Raymond 1985a [0.2--4\,keV]; 
Eracleous et al.\ 1991a [0.1--3.5\,keV; 2--10\,keV]; 
Mukai \&\ Shiokawa 1993 [2--10\,keV]; Richman 1996 [0.1--2.4\,keV]).
This translates to, generally, $L_{\rm X}$$\simeq$10$^{29}$--10$^{32}$\,erg\,s$^{-1}$.
All the CVs below 10$^{30}$\,erg\,s$^{-1}$ are short-period DNe
or low-state magnetic CVs (Verbunt et al.\ 1997).

Among the non-magnetic CVs
the ratio of the X-ray flux to optical and/or UV flux,
$F_{\rm X}$/$F_{\rm opt}$, decreases along the sequence SU\,UMa stars 
($F_{\rm X}$/$F_{\rm opt}$$\sim$0.1) -- U\,Gem stars -- Z\,Cam stars 
($F_{\rm X}$/$F_{\rm opt}$$\sim$0.01) -- UX\,UMa stars
($F_{\rm X}$/$F_{\rm opt}$$\lesssim$10$^{-3}$), due mainly to variations in the optical/UV flux
(Verbunt et al.\ 1997; see also van Teeseling \&\ Verbunt 1994; 
van Teeseling et al.\ 1996).
Of the few exceptions are, e.g., the double degenerate AM\,CVn systems (Sect.~\ref{amcvn}).
We note that for magnetic CVs $F_{\rm X}$/$F_{\rm opt}$ is comparable to
that of the SU\,UMa stars (Verbunt et al.\ 1997).
The general pattern is in agreement with Patterson \&\ Raymond (1985a; 
see also Richman 1996) 
who find that non-magnetic CVs with high $\dot{M}$ show low $F_{\rm X}$/$F_{\rm opt}$, 
and in agreement with the fact that $F_{\rm X}$/$F_{\rm opt}$
is seen to decrease with increasing $P_{\rm orb}$
(van Teeseling \&\ Verbunt 1994; van Teeseling et al.\ 1996; see also
C\'ordova \&\ Mason 1984)\footnote{We caution, however, that the measurements for 
CVs with short orbital periods are biased
towards quiescent systems (i.e., CVs wit low $\dot{M}$). X-ray observations of high $\dot{M}$ short 
orbital period CVs could resolve the issue.}. 
This latter correlation stems from the fact that the UV flux is a 
strongly increasing function of $P_{\rm orb}$, which 
in turn is likely related to $\dot{M}$: a high $\dot{M}$ apparently causes the disk 
to emit more UV flux, but not more X-ray flux (see, e.g., van Teeseling et al.\ 1996).
This general pattern is somewhat perturbed, however, 
by the anti-correlation between the inclination, $i$,
and the observed X-ray flux (van Teeseling et al.\ 1996;
see also Patterson \&\ Raymond 1985a).

An empirical relation between the equivalent width, EW,
of the optical H$\beta$ emission line and $F_{\rm X}$/$F_{\rm opt}$ exists
(Patterson \&\ Raymond 1985a; Richman 1996). 
This relation predicts $F_{\rm X}$/$F_{\rm opt}$ to within a factor of 3.
The correlation of $F_{\rm X}$/$F_{\rm opt}$ with EW(H$\beta$) is thought 
to reflect also an underlying correlation with $\dot{M}$, 
since EW(H$\beta$) is known to correlate with the absolute visual magnitude
of the disk, which in turn is correlated with $\dot{M}$ (e.g, Patterson 1984; 
Warner 1995). Thus low $\dot{M}$ systems produce strong H$\beta$ emission lines and a larger
$F_{\rm X}$/$F_{\rm opt}$. 

The absorption column densities, $N_{\rm H}$, as derived from X-ray spectral fits
are generally in the range 10$^{20}$--10$^{21}$\,cm$^{-2}$ 
(e.g., Eracleous et al.\ 1991a; Richman 1996).
VW\,Hyi has one of the lowest values of $N_{\rm H}$ for any CV
(6$\times$10$^{-17}$\,cm$^{-2}$; Polidan et al.\ 1990), 
which makes it an ideal CV to study in the EUV and soft X-ray range. 
Comparison of the $N_{\rm H}$ values with the
colour excess $E_{B-V}$ derived from the 2200\AA\ feature
(Verbunt 1987), 
shows that $N_{\rm H}$ is often higher than predicted 
on the basis of the average relation derived by Predehl \&\ Schmitt (1995).
The excess column may be related to absorbing gas in the CV itself, which
can be responsible for some of the orbital variations seen, especially
since the CVs displaying high absorbed spectra are known to have a high 
inclination (e.g., Eracleous et al.\ 1991a; Verbunt 1996; 
see also Sects.~\ref{quiescence}, \ref{outburst}).

The X-ray spectral flux distributions within the 0.5--2.5\,keV band 
seem to be fairly similar for most CVs (Verbunt et al.\ 1997),
although it appears that VY\,Scl stars in their high state
and non-SU\,UMa stars  have somewhat harder spectra
than SU\,UMa and UX\,UMa stars (van Teeseling et al.\ 1996).
Individual systems, however, may 
show significant epoch-to-epoch variability both in luminosity
and temperature (e.g., Mukai \&\ \mbox{Shiokawa} 1993). 
X-ray spectra of non-magnetic CVs generally show (bremsstrahlung) 
temperatures in the range 1--5\,keV
(e.g., C\'ordova et al.\ 1981; 
C\'ordova \&\ Mason 1983, 1984;
Eracleous et al.\ 1991a; 
Mukai \&\ Shiokawa 1993).
Quiescent DNe are hard X-ray sources with
(bremsstrahlung) temperatures of a few keV
(e.g., Patterson \&\ Raymond 1985a;
Vrtilek et al.\ 1994) 
up to $\sim$10\,keV 
(e.g., C\'ordova \&\ Mason 1983). 
Systematic residuals in the fitted X-ray spectra, however, already indicated that the X-ray emission
is not well described by the single-temperature models,
both for low and high $\dot{M}$ CVs,
but instead must be described by a range of temperatures (e.g.\ 
Eracleous et al.\ 1991a;
Richman 1996; see Sect.~\ref{spectra}).

The emission measure, EM\footnote{Defined as EM=$\int\! n^2_{\rm e}dV$, 
where $n_{\rm e}$ is the electron density and $V$ the emitting volume.}, 
is not a strong function of $\dot{M}$ 
(van Teeseling et al.\ 1996). 
It decreases, however, for CVs with higher $i$.
For CVs with $i$$<$70$^{\circ}$ this cannot be due to obscuration 
of the X-ray source by matter in the outer parts of the disk or by the 
donor. The anti-correlation between EM and $i$
excludes models in which the X-rays are emitted in a relatively large 
optically thin volume. On the other hand, if the X-rays originate 
from the inner part of the disk 
and the scale height of the optically thin 
X-ray source is not much higher than the disk thickness
(as eclipse observations of quiescent DN suggest, see Sect.~\ref{quiescence}), EM could depend on $i$.
Van Teeseling et al.\ (1996) come to the conclusion that in 
high-inclination systems most of the X-ray flux is absorbed by the disk.

So, what is the origin of the X-ray emission in non-magnetic CVs?
In the next subsection we describe the major source of X-rays in these systems,
the BL.

\subsection{The boundary layer model}
\label{bl}

In non-magnetic CVs the accretion is governed by the disk. 
Basic theory predicts that half of the gravitational
potential energy of the accreting material is liberated through the viscosity
in the disk, while the other half is liberated in a boundary layer (BL) between
the disk and the surface layer of the WD 
(e.g., Shakura \&\ Sunyaev 1973; 
Lynden-Bell \&\ Pringle 1974; 
Pringle 1981). 
Material in the BL moves with Keplerian speeds 
and collides with the WD which is presumably rotating more slowly than break-up velocity.
This results in luminosities of the disk and BL of
$L_{\rm disk}$$\simeq$$L_{\rm BL}$$\simeq$$GM_{\rm WD}\dot{M}/2R_{\rm WD}$,
where $M_{\rm WD}$ and $R_{\rm WD}$
are the mass and radius of the WD, respectively.
For a WD with $M_{\rm WD}$=1\,M$_{\odot}$, $R_{\rm WD}$=$10^9$\,cm
and $\dot{M}$=10$^{-10}$\,M$_{\odot}$\,yr$^{-1}$ this amounts
to about 4$\times$10$^{32}$\,erg\,s$^{-1}$.
With these modest luminosities the X-rays do not strongly influence
the appearance of the disk, as they do in bright LMXBs.
The disk is generally too cool ($kT$$<$1\,eV) to emit X-rays.
It radiates mostly at optical and ultraviolet (UV) wavelengths,
whereas the BL mostly radiates in the extreme ultraviolet (EUV) and in X-rays
(e.g., Bath et al.\ 1974b). 

When $\dot{M}$ is low, such as in DNe in quiescence,
the BL is observed to be optically thin. Shocks make the gas hot with a temperature
of about the virial temperature, $kT_{\rm vir}$=$GM_{\rm WD}m_{\rm H}/6kR_{\rm WD}$$\sim$20\,keV.
When $\dot{M}$ is high, such as in DNe in outburst or in NLs, 
the BL is observed to be optically thick. Cooling of the BL is efficient
and the X-ray spectrum is thermalised with an approximate black-body temperature of
$kT_{\rm bb}$=$(GM_{\rm WD}\dot{M}/8\pi\sigma R^3_{\rm WD})^{1/4}$$\sim$10\,eV
and $L_{\rm X}$$>$10$^{34}$\,erg\,s$^{-1}$
(e.g., Pringle 1977; 
Pringle \&\ Savonije 1979; 
Tylenda 1981; 
Narayan \&\ Popham 1993, 
Popham \&\ Narayan 1995).
The critical $\dot{M}$, $\dot{M}_{\rm crit}$, 
generally depends on $M_{\rm WD}$ and the viscosity in the disk,
and is about 10$^{-10}$\,M$_{\odot}$\,yr$^{-1}$.

Although X-ray observations of CVs are often interpreted within the above
described framework of the standard BL model
(e.g., many quiescent DNe were found to be modest hard X-ray sources), 
evidence that BLs really exist in CVs is generally indirect.
In the next subsections we provide the observational efforts 
to find this BL, both in quiescence and outburst of DNe, as well as in other
high-$\dot{M}$ CVs.

\subsection{Quiescent dwarf novae}
\label{quiescence}

For CVs with $\dot{M}$$<$$\dot{M}_{\rm crit}$ 
Patterson \&\ Raymond (1985a) 
show that the hard X-ray data from {\it Einstein} were generally consistent with hot optically 
thin emission from the BL.
The observed temperatures are in the range expected.
However, most of these CVs show less BL radiation than predicted
(e.g., Pringle et al.\ 1987; Belloni et al.\ 1991; 
van Teeseling \&\ Verbunt 1994;
Vrtilek et al.\ 1994). 
This also holds for high $\dot{M}$ systems (see Sect.~\ref{outburst}), and it 
is referred to as `the mystery of the missing BL' (see Ferland et al.\ 1982). 
Different explanations for this lack in quiescence have been suggested:
disruptions of the inner disk by magnetic fields (Livio \&\ Pringle 1992; 
Lasota et al.\ 1995; Warner et al.\ 1996), 
coronal siphon flows (Meyer \&\ Meyer-Hofmeister 1994; Lasota et al.\ 1995), 
or irradiation by the (relatively) hot WD (King 1997); 
a rapidly rotating WD (e.g., Ponman et al.\ 1995);
a rapidly spinning accretion belt (Sion et al.\ 1996);
reflection effects and cooling flows (Done \&\ Osborne 1997). 
Alternatively, the BL largely radiates an additional
{\em very} soft component ($kT$$\lesssim$10\,eV) which would
remain undetectable due to interstellar absorption
(e.g., Patterson \&\ Raymond 1985b). 

High-inclination CVs provide an opportunity to locate the X-ray emitting regions.
When the X-ray source is eclipsed one can constrain its size and
location using the orbital phase, $\phi_{\rm orb}$, and the duration of ingress and egress.
One factor complicating such observations, however, is that
eclipsing CVs tend to be fainter in X-rays than low-inclination
CVs, so eclipse studies have been
rather count-rate limited (e.g., van Teeseling et al.\ 1996).

X-ray eclipses have been seen during quiescence of the DNe
HT\,Cas (Mukai et al.\ 1997; $i$$\simeq$81$^{\circ}$), 
Z\,Cha (van Teeseling 1997a; $i$$\simeq$82$^{\circ}$) and 
OY\,Car (Pratt et al.\ 1999a; 
Ramsay et al.\ 2001a; 
Wheatley \&\ West 2002; 
$i$$\simeq$83$^{\circ}$), as well as during a low state in quiescence 
of HT\,Cas (Wood et al.\ 1995a). 
They all occur at the time of the optical WD eclipse.
The X-ray ingress and egress are rapid; 
in OY\,Car their duration is significantly shorter
than in the optical (30$\pm$3\,s vs.\ 43$\pm$2\,s; Wheatley \&\ West 2002). 
This is consistent with the indication that the total X-ray eclipse in HT\,Cas
has a slightly shorter duration
than the optical one (by about 27\,sec; Mukai et al.\ 1997). 
If one assumes that the optical contact points represent the 
contact points of the WD, then the X-ray emitting
region must be smaller than the WD, possibly originating
from a broad equatorial belt of which the lower
half is absorbed (Mukai et al.\ 1997; van Teeseling 1997a; Ramsey et al.\ 2000a; Wheatley \&\ West 2002).
This also explains the observed anti-correlation between $i$ and
the EM (van Teeseling et al.\ 1996; see Sect.~\ref{overview}).
Since the WD is too cool to produce X-rays, a BL
must be responsible for the out-of-eclipse X-ray emission (see Mukai et al.\ 1997). 

In eclipse there is residual X-ray emission at about
1\%\ of the out-of-eclipse flux, with $L_{\rm X}$$\sim$3$\times$10$^{28}$\,erg\,s$^{-1}$
and a soft spectrum ($kT$$\sim$1\,keV).
The fact that this emission in eclipse is softer than that seen
out of eclipse seems to rule out the possibility that the residual flux
is BL emission scattered into our line of sight by
circumstellar material (Wheatley \& West 2002).
The luminosity (e.g., Rosner et al.\ 1985; Hempelmann et al.\ 1995), 
as well as the temperature (e.g., Schmitt et al.\ 1990), 
are consistent with coronal emission from a cool main-sequence donor.
Ramsay et al.\ (2001a), however, argue that the residual emission may come
from a weak remnant of a large corona, 
which is more prominent during outburst
(see Sect.~\ref{outburst}).

Apart from eclipses, 
dips in the X-ray light curves have been observed, up to $\sim$50\%\ deep, during quiescence
in U\,Gem (at $\phi_{\rm orb}$$\sim$0.3 and 0.8: 
Szkody et al.\ 1996, 2000a; $i$$\sim$65$^{\circ}$),
Z\,Cha ($\phi_{\rm orb}$$\sim$0.7--0.8: van Teeseling 1997a),
WZ\,Sge ($\phi_{\rm orb}$$\sim$0.7: Patterson et al.\ 1998; $i$$\sim$75$^{\circ}$),
and OY\,Car ($\phi_{\rm orb}$$\sim$0.2--0.5: Ramsay et al.\ 2001a). 
They are only apparent at low X-ray energies, which indicates absorption 
effects. Similar kinds of dips have also been found during outburst
(see Sect.~\ref{outburst}).

In order for the dips to be visible at inclinations such as in U\,Gem
the material must be located far from the orbital plane
(e.g., Mason et al.\ 1988; Naylor \&\ La Dous 1997).
Note that in quiescence the X-ray dips in U\,Gem were less deep than
during outburst. This means that the X-ray emitting region must be only
slightly larger than the outburst BL and that the absorbing
material that was present at outburst must maintain a similar location in
quiescence.
The small residual X-ray flux seen may be scattered into the line of
sight from high above the plane (by a disk corona or a wind;
e.g., Naylor \&\ La Dous 1997; Mason et al.\ 1997; see also Sect.~\ref{spectra})
or may possibly originate from a hot corona of the donor (Wood et al.\ 1995b).
{\it HST} observations of OY\,Car in quiescence
show that the UV emission from the WD surroundings
is also absorbed by matter above the disk, which is referred to as an 
`iron curtain' (see Horne et al.\ 1994). 
However, this curtain does not always seem to exist
(Pratt et al.\ 1999a). 
Note that no dips were found in the quiescent UV light curves of OY\,Car 
(Ramsay et al.\ 2001a).

Similar kinds of absorption dips have also been seen in IPs (see Sect.~\ref{IPs}),
as well as in LMXBs (e.g., Mason 1986;
Parmar \&\ White 1988; 
White et al.\ 1995; 
Kuulkers et al.\ 1998, and references therein).
Note that column densities of
$\lesssim$10$^{19}$\,cm$^{-2}$ to $\sim$10$^{22}$\,cm$^{-2}$ in CVs
(e.g., Naylor \&\ La Dous 1997) are sufficient to extinguish soft
X-ray emission during the dips, whereas maximum column densities of
$>$10$^{23}$\,cm$^{-2}$ are typically recorded in LMXBs.

A popular model for the dips is the one outlined by Frank et al.\ (1987). 
They explain the dips as the interaction of the accretion stream with the disk, which
splashes material out of the plane to form cool clouds that obscure the radiation produced
close to the compact object
(see also Armitage \&\ Livio 1996, 1998; 
Kunze et al.\ 2001). 
For CVs they predict a single broad dip between phases 0.6 and 0.8, exactly as observed in, e.g., U\,Gem.

\subsection{Outbursting dwarf novae and other high-$\dot{M}$ CVs}
\label{outburst}

For most CVs in quiescence $\dot{M}$ onto the WD is
of the order of 10$^{-12}$--10$^{-11}$\,M$_{\odot}$\,yr$^{-1}$
(e.g., Patterson 1984; Warner 1995). 
During an outburst $\dot{M}$ increases by $\sim$2 orders of magnitude, so the disk is
likely to cross $\dot{M}_{\rm crit}$. 
The CV is then expected to change from a hard to a soft X-ray emitter.
However, the situation appears to be not that simple, as we will show below.

Soft X-rays have been detected during outbursts
of SS\,Cyg (e.g., Rappaport et al.\ 1974; Mason et al.\ 1978; 
C\'ordova et al.\ 1980b; 
Jones \&\ Watson 1992;
Ponman et al.\ 1995),
U\,Gem (e.g., C\'ordova et al.\ 1984), 
VW\,Hyi (van der Woerd et al.\ 1986; 
Mauche et al.\ 1991; 
van Teeseling et al.\ 1993; 
Wheatley et al.\ 1996b), 
SW\,UMa (Szkody et al.\ 1988), 
and Z\,Cam 
(Wheatley et al.\ 1996a). 
Other high-$\dot{M}$ CVs generally do not show the soft component 
(e.g., Silber et al.\ 1994; 
van Teeseling et al.\ 1995).
When a soft component is present the X-ray spectra show $kT_{\rm bb}$$\sim$5--30\,eV;
these temperatures are similar to the BL temperatures derived from 
high resolution EUV and X-ray spectra (Mauche et al.\ 1995: SS\,Cyg; 
Long et al.\ 1996: U\,Gem; 
Mauche 1996b: VW\,Hyi; 
Mauche \&\ Raymond 2000: OY\,Car; see also Sect.~\ref{spectra}).
Note that not all of the soft component
is optically thick (Mauche et al.\ 1995; 
Long et al.\ 1996). 
The soft X-ray fluxes increase by a factor of $\sim$100 from 
quiescence to outburst. However, they are still too low compared to the 
simple BL models (e.g., Mauche et al.\ 1991; 
van Teeseling et al.\ 1993; 
van Teeseling \&\ Verbunt 1994; 
Ponman et al.\ 1995;
Wheatley et al.\ 1996b), 
similar to the discrepancy seen in quiescence (see Sect.~\ref{quiescence}).
A study of the ionization states inferred from the P\,Cygni
lines arising in winds from high $\dot{M}$ CVs led 
to a similar conclusion
(Drew \&\ Verbunt 1985; 
Hoare \&\ Drew 1991). 
Various explanations for the discrepancy (or absence) 
of soft X-ray flux during outburst have been put forward:
differences in $N_{\rm H}$ to different systems (e.g., Patterson \&\ Raymond 1985b;
Long et al.\ 1996); 
differences in $M_{\rm WD}$ and the WD rotation (see below); 
absorption in the disk wind (Jensen 1984; 
Kallman \&\ Jensen 1985); 
energy loss in the form of a wind (e.g., Silber et al.\ 1994;
Ponman et al.\ 1995). 
Moreover, changes in the BL temperature 
can shift most of the flux out of the soft X-ray bandpass 
(e.g., C\'ordova et al.\ 1980a; 
Patterson \&\ Raymond 1985b). 

\begin{figure}[t]
\hspace{-1.0cm}
\resizebox{0.50\hsize}{!}
{\includegraphics[clip=]{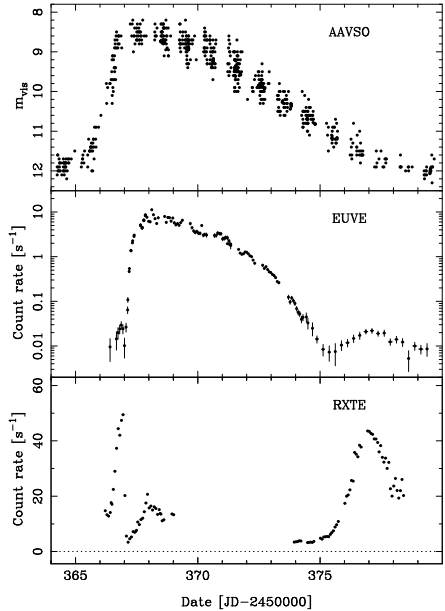}}
\hfill
\parbox[b]{0.52\hsize}{
\caption{\label{sscyg} 
Simultaneous AAVSO, {\it EUVE} and {\it RXTE} observations of SS\,Cyg throughout
outburst. Note that the {\it RXTE} light curve is plotted on a linear scale
in order to emphasize the timing of the sharp transitions. From Wheatley et al.\ (2003).
\vspace{0.5cm}
}
}
\end{figure}

Our knowledge of the evolution of the spectral flux distribution 
during outbursts of DNe at various wavelengths is mainly based
on fragmented (nearly) simultaneous observations.
A few dedicated campaigns do exist, however
(see, e.g., Pringle et al.\ 1987; Wheatley et al.\ 1996b; 
Szkody 1999, and references therein).
One of the most complete coverages to date of a DN outburst, 
is that of SS\,Cyg (Mauche \&\ Robinson 2001; Wheatley et al.\ 2000; 
see Fig.~\ref{sscyg}).  
We here describe the general behaviour seen at EUV and X-ray wavelengths.

The soft X-rays lag the optical outburst light curve by about 12--36\,h during the
rise (e.g., Jones \& Watson 1992; Mauche \&\ Robinson 2001). 
This is comparable to that measured in the 
far-UV ($\lesssim$10\,eV; Polidan \&\ Holberg 1984). 
Wheatley et al.\ (2000) found that the X-ray outburst of SS\,Cyg 
started $\sim$18\,h {\it before}
the EUV one.
The start of the X-ray outburst is marked by a sudden softening of the 
X-ray spectrum; the rise to soft X-ray maximum is rapid
(e.g., Wheatley et al.\ 2000).
After reaching maximum early in the outburst, it rapidly decreases again
(but less fast than the rise).
The decrease is more rapid towards shorter wavelengths.
The soft X-rays lead the optical light curve during the decline, 
and disappear before the end of the optical outburst 
(e.g., van der Woerd et al.\ 1986; Mauche \&\ Robinson 2001). 
The soft X-ray rise and decay times are shorter with respect to the optical
(e.g., van der Woerd et al.\ 1986;
Jones \&\ Watson 1992).

The initial soft X-ray rise could be the arrival of the heating wave
through the disk at the BL, and the sudden spectral softening is 
as expected in the BL models. The rapid rise time may represent  
the time scale of the transition between optically thin and thick emission;
the less rapid drop at the end may
represent the time scale of the 
inverse process (e.g., Jones \&\ Watson 1992; Wheatley et al.\ 2000). 

At the time the soft X-rays appear, the hard X-ray flux is suppressed
(e.g., Wheatley et al.\ 2000; Baskill et al.\ 2001). They do not disappear, however. 
They stay present during the outburst, with somewhat lower
temperature and flux than in quiescence.
This may be attributed to a density gradient in the optically thick
BL, such that there is always a hot optically thin layer
which emits hard X-rays (e.g., Patterson \&\ Raymond 1985a;  
Done \&\ Osborne 1997).
The anti-correlation between the soft and hard X-ray flux
suggests that we see two physically distinct emission components.
The coincidence in the timing show that they are related, however,
and possibly mark the time at which the BL becomes optically thick.
The hard X-ray flux during outburst is considerably smaller than the soft X-ray flux 
(e.g., $\sim$0.1\%\ during 
a superoutburst of VW\,Hyi; van der Woerd et al.\ 1986) 
and it declines throughout the outburst (e.g.\ Verbunt et al.\ 1999).
The temperature of the hard X-ray component 
increases from outburst to quiescence (e.g., Hartmann et al.\ 1999). 
The decline in flux and increase in temperature probably reflects the 
(slowly) decreasing $\dot{M}$ on the WD
(e.g., Jones \&\ Watson 1992; Hartmann et al.\ 1999).

The hard X-ray flux recovers to quiescent levels
just at the very end of the optical outburst
(Wheatley et al.\ 1996b, 2000; see also Yoshida et al.\ 1992; 
van Teeseling \&\ Verbunt 1994; Ponman et al.\ 1995). 
The recovery time scale is slightly longer than that of the optical decline
(e.g., Jones \&\ Watson 1992).
The hard X-ray flux varies on a time scale of hundredths of seconds with an amplitude of $\sim$100\%\
at the start of the recovery to 50\%\ at the end of that observation, with
the hardness ratio staying constant.
As the end of the optical outburst is thought to correspond to the cooling 
of the disk region immediately surrounding the WD, this observation
indicates that the hard X-rays originate from an area of the disk very close to the WD 
(e.g., Wheatley et al.\ 1996b; Verbunt 1996). 
Note that at the time of the hard X-ray recovery the EUV light curve of 
SS\,Cyg exhibited a secondary maximum, which emission is consistent with
the soft tail of the hard X-ray emission (Mauche \&\ Robinson 2001; 
Wheatley et al.\ 2000, 2002). 

The situation is different for the outbursts of U\,Gem, where
both soft and hard X-ray fluxes are higher 
during outburst than in quiescence (by a factor of $\sim$10--100).
The largest increase occurs at EUV wavelengths.
This corresponds to the expected increase
in an optically thick BL radiating at temperatures near 
10\,eV at outburst (e.g., Szkody et al.\ 1999). 
While the optical flux stays
constant near maximum, the EUV flux drops 
(Long et al.\ 1996). 
Assuming the EUV flux originates from near the WD, this suggests 
that $\dot{M}$ in the innermost regions of the disk decreases
compared to the outer regions (which presumably are still optically thick).
During the outburst decline, both the soft and hard X-rays decrease
faster than the optical flux
(Mason et al.\ 1978; 
Swank et al.\ 1978; 
C\'ordova \&\ Mason 1984; 
see also Szkody et al.\ 1999). 
Values derived for $L_{\rm BL}$/$L_{\rm disk}$ are
$\sim$0.5 during quiescence (Szkody et al.\ 1999), 
which is among the highest for DNe, and $\sim$1 during outburst
(Long et al.\ 1996).
So, in quiescence, as at outburst, U\,Gem comes closest to the standard BL model.

There are several reasons why $L_{\rm BL}$ may be 
larger in U\,Gem than in systems like VW\,Hyi 
(see Long et al.\ 1996, and references therein).
Studies (Pringle 1977; Popham \&\ Narayan 1995)
have shown that, when $\dot{M}$ is held fixed, the BL temperature
and $L_{\rm BL}$ increase substantially with WD mass.
On the other hand, rotation of the WD decreases both the amount of energy 
released and the effective temperature of the BL.
The WD in U\,Gem is more massive (1.0--1.2\,M$_{\odot}$)
than WDs in most DNe and VW\,Hyi in particular 
($\sim$\,0.6\,M$_{\odot}$). It
appears to be at most slowly rotating
($v\sin{i}$$\lesssim$100\,km\,s$^{-1}$), while the WD in 
VW\,Hyi rotates with a $v\sin{i}$$\simeq$400\,km\,s$^{-1}$
(e.g., Sion et al.\ 2002), 
which corresponds to 20\%\ of the break-up velocity. 

In contrast to quiescence, there are no eclipses in the X-ray and EUV light curves 
of OY\,Car during outburst
(Naylor et al.\ 1988; 
Pratt et al.\ 1999b; 
Mauche \&\ Raymond 2000), 
as well as in the NL UX\,UMa (Wood et al.\ 1995b). 
This suggests that the prime X-ray source, probably the BL, is obscured
at all orbital phases.
From contemporaneous observations at other wavelengths 
(Naylor et al.\ 1987, 1988), 
extensive azimuthal structures on the outer disk
had been inferred, which may block our view of the BL region. 
The observed dips in various other CVs (see Sect.~\ref{quiescence} and below) are also explained this way.
The X-rays we see are thought to be emitted or scattered by a more 
extended source (e.g., Verbunt 1996); 
e.g., due to a disk corona (Naylor et al.\ 1988) or
scattering from a photo-ionised disk wind 
(Raymond \&\ Mauche 1991; Mauche \&\ Raymond 2000; see also Sect.~\ref{spectra}). 

Dips in the EUV and X-ray light curves during outburst have been seen in U\,Gem 
during a normal outburst (at $\phi_{\rm orb}$$\sim$0.8; Long et al.\ 1996),
as well as during an anomalously long ($\sim$45~days) outburst
(Mason et al.\ 1988). As in quiescence
(Sect.~\ref{quiescence}), the dips only occur at low energies, indicating
absorption effects.
The morphology of the dips changes from cycle to cycle, 
related to changes in the absorbing material.
The dips are deeper at shorter wavelengths, 
suggesting that the hot central area around the WD 
is being obscured by cooler material further out.
During one dip observed by Mason et al.\ (1988) the X-ray source was
completely extinguished in 15\,s, putting the absorbing material
near the outer edge of the disk.

\subsection{X-ray spectral features}
\label{spectra}

Most of the surveys done so far showed that single and sometimes two-temperature
bremsstrahlung models were sufficient to describe the CV X-ray spectra
(see Sect.~\ref{overview}), except for the occasional inclusion of a Gaussian to 
represent a line near 6.7\,keV (see below). This is mainly due either to rather poor energy
resolution and/or poor statistics. The use of
more realistic models was generally not warranted.
With the advent of better resolution, larger collecting area, better
photon-counting devices, broader band passes, it became
clear that the spectra are far more complicated.
The X-ray spectra of DNe and NLs can probably be best decribed as somewhere between 
a pure bremsstrahlung model and a pure coronal model (e.g., 
van Teeseling \&\ Verbunt 1994).
Generally, X-ray spectra from non-magnetic CVs are due to hot thermal plasma
in the BL, even at high $\dot{M}$ (e.g., Mukai 2000).
This is because the shock-heated plasma in the BL must cool from a 
temperature near 10\,keV indicated by the X-ray spectra to the 
photospheric temperature of the WD
($\sim$2.5\,eV). 
The situation may be further complicated since X-rays from the 
hottest gas can photoionize cooler gas, altering both the 
energy balance and the ionization state at intermediate temperatures.

Many CVs, either in quiescence or in outburst,
show an emission line near 6.7\,keV from the 
K$\alpha$ transition of highly ionised Fe, with EW$\sim$0.8--1.0
(e.g., Szkody et al.\ 1990); 
it is associated with the hard X-ray emitting, optically thin plasma.
Line emission near 7.9\,keV has been reported just after an outburst of
SS\,Cyg (Jones \&\ Watson 1992) and OY\,Car (Ramsay et al.\ 2001b).
This may be interpreted as thermal Fe-K$\beta$ emission, 
confirming the origin of line emission from a hot optically thin region.
The presence of an absorption edge near 8.3\,keV in the {\it Ginga} spectrum
of SS\,Cyg implies substantial covering of the hard
X-ray emission by the highly ionised gas (possibly a wind). 
Note that this is hard to reconcile
with the picture in which the hard X-rays arise from a hot corona
(Yoshida et al.\ 1992). 
SS\,Cyg also shows a reflection component both in quiescence and outburst
(Done \&\ Osborne 1997). 
Its contribution is larger in the softer X-ray spectra 
seen in outburst than in quiescence.
This supports models in which the quiescent inner disk is not present
or not optically thick, so that the only reflector is the WD surface
rather than the WD plus disk. The amount of reflection in outburst
is also more consistent with the hard X-rays forming a corona over the 
WD surface rather than just an equatorial belt as seen in quiescence.
Note that a reflection component is absent in OY\,Car; this possibly is due 
to the high inclination, so that it may be obscured by the disk 
(Ramsay et al.\ 2001b).

Although {\it ASCA} showed complex structures in the 
X-ray spectra, the spectral resolution was still not high enough
to resolve individual lines, especially in regions where a lot
of lines are expected (e.g., Fe\,L complex around 1\,keV). 
First {\it EUVE}, and now {\it Chandra} and {\it XMM-Newton} provide the 
opportunity to perform detailed temperature 
diagnostics from individually resolved lines and line ratios.
Line ratios can be used to constrain the electron density, electron temperature,
and ionization balance (see, e.g., Mauche et al.\ 2001; 
Szkody et al.\ 2002a, and references therein).
They also provide enough velocity resolution to begin to study the
effects of velocity broadening, which gives important clues to whether the emitting
region is located in a rapidly rotating BL or is close to the more slowly
rotating WD.

High resolution X-ray spectra of U\,Gem in quiescence (Szkody et al.\ 2002a) 
revealed prominent narrow emission lines of O, Ne, Mg, Si, S, and Fe. 
The line fluxes, ratios, and widths
indicate that the X-ray emission lines arise from a range of temperatures in 
a high density ($>$10$^{14}$\,cm$^{-3}$) gas, moving at low
($<$300\,km\,s$^{-1}$) velocity, with a small ($<$10$^{7}$\,cm) scale height compared
to the WD radius. This is consistent with a BL, as was also inferred from the eclipse
light curves (see Sect.~\ref{quiescence}). 

\begin{figure}[t]
\hspace{-1.2cm}
\resizebox{0.80\hsize}{!}
{\includegraphics[angle=-90.0,bbllx=48pt,bblly=36pt,bburx=579pt,bbury=738pt,clip=yes]{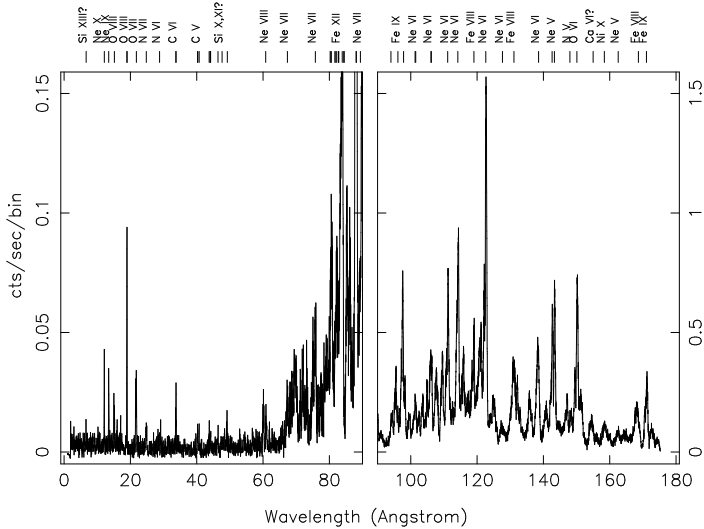}}
\hfill
\parbox[b]{0.24\hsize}{
\caption{\label{letgs} 
Average Chandra/LETGS spectrum 3.5\,d into the superoutburst of WZ\,Sge. 
Note the difference in scale between the short and long-wavelength part of the spectrum. 
Indicated are many of the line identifications. From Kuulkers et al.\ (2002).
\vspace{-6.55cm}
}
}
\end{figure}

The EUV/soft X-ray spectra of U\,Gem in outburst ($\phi_{\rm orb}$$\sim$0.6--0.8, Long et al.\ 1996), 
and OY\,Car (Mauche \&\ Raymond 2000) and WZ\,Sge (Kuulkers et al.\ 2002; Fig.~\ref{letgs}) 
in superoutburst resemble each other markedly, and are unlike that 
seen in other DNe, which is attributed to their high inclination.
Below $\sim$0.2\,keV ($\gtrsim$65\AA\/) they show a `forest' of broad 
(FWHM$\sim$800--1200\,km\,s$^{-1}$) emission lines of intermediate 
ionization stages of N, O, Ne, Mg and Fe, on top of a continuum
(which is weak in OY\,Car and WZ\,Sge, and appears to be line-free in U\,Gem).
The phase resolved spectra of U\,Gem show that the eclipses affect the 
continuum more strongly than the lines, implying
that the lines are produced in a region of larger extent than that
of the continuum, which is presumably formed in the BL.
The line identifications alone significantly constrain the physical
nature of the emitting plasma. 
Because all of the above strong lines are resonance lines, 
good spectral fits are obtained with a model
wherein the radiation from the BL and disk is scattered
into the line of sight by the system's photo-ionised disk wind
(Mauche \&\ Raymond 2000).
Note that the absence of X-ray eclipses in OY\,Car during superoutburst
(Naylor et al.\ 1988) can be understood if much of the X-ray emission we see
is light scattered in such an extended wind.

The EUV lines arise from the dominant ionization states of the 
wind, and their strengths suggest that the wind mass-loss rate in U\,Gem is 
a substantial fraction of the WD accretion rate (Long et al.\ 1996).
The lines are effectively the emission peaks of P\,Cygni profiles. This
requires the scattering region to be of order
10$^{10}$\,cm, similar to the size of the region inferred from 
studies of the UV P\,Cygni lines in other
DNe and NLs. These P\,Cygni profiles have been shown 
to arise in strong winds with terminal velocities of order
3000\,km\,s$^{-1}$ (e.g., Drew \&\ Verbunt 1985; 
Mauche \&\ Raymond 1987; 
Woods et al.\ 1990). 

WZ\,Sge's spectrum at higher energies, $\gtrsim$0.2\,keV
($\lesssim$65\AA\/), 
shows weaker lines of highly ionised ions, with
O\,{\sc VIII} (0.66\,keV; 18.9\AA\/) standing out.
A strong emission line was also seen near 2.4\,keV,
probably associated with He-like S\,{\sc XV}, which cannot be accounted for 
using an optically thin thermal plasma model
(Kuulkers et al.\ 2002).

Because of the recent access to high quality X-ray spectra, `new' interpretations are also 
emerging. This is demonstrated by, e.g., Mukai et al.\ (2003), who show that the {\it Chandra} data
of the DNe SS\,Cyg and U\,Gem and the old nova V603\,Aql are well described by 
a simple cooling-flow model. This in contrast to three IPs which are better
described by a photo-ionization model. 

\subsection{VY\,Scl stars}
\label{vyscl}

During optical high states NLs generally have similar X-ray spectral
characteristics, with temperatures between $kT_{\rm bb}$$\sim$0.25--0.5\,keV
and $L_{\rm X}$$\sim$10$^{31}$--10$^{32}$\,erg\,s$^{-1}$.
Since VY\,Scl stars are thought to have a high $\dot{M}$ during the high state
their X-ray spectra should be rather similar to
DNe in outburst. However, this seems not to be the case.
Distinct differences at other wavelengths exist as well; 
see Greiner (1999) 
for an extensive comparison between VY\,Scl stars and DNe in outburst.

A very soft component ($kT_{\rm bb}$$\sim$19\,eV) was found
in V751\,Cyg during an optical low state, at
which time the bolometric X-ray luminosity was around 20 times higher 
($L_{\rm X}$$\simeq$5$\times$10$^{36}$\,erg\,s$^{-1}$) than in the high state
(Greiner et al.\ 1999; but see Patterson et al.\ 2001). 
V\,Sge showed a similar behaviour: it is a faint hard X-ray source 
during optical bright states, while during 
optical low states it shows X-ray luminosities similar to V751\,Cyg 
(Greiner \&\ van Teeseling 1998). 
$L_{\rm X}$ is clearly higher than generally observed
from CVs (see Sect.~\ref{intro}), 
and compatible with the lower end of the luminosity distribution
of supersoft sources (SSS; see Chapter 11).
It was therefore suggested that VY\,Scl stars in their low states may have a link with
SSS (Greiner et al.\ 1999). 
Not all VY\,Scl stars in their low state show high values of
$L_{\rm X}$, however.
Examples are KR\,Aur with $L_{\rm X}$$\sim$10$^{31}$\,erg\,s$^{-1}$
(Eracleous et al.\ 1991a;
Schlegel \&\ Singh 1995) 
and MV\,Lyr with $L_{\rm X}$$\lesssim$5$\times$10$^{29}$\,erg\,s$^{-1}$
(Greiner 1999) during a low state.

The X-ray spectra during a
high (KR\,Aur) and intermediate (TT\,Ari) optical state were shown, however, to be poorly 
described by black-body radiation; a thermal plasma model described the data better.
It was concluded, therefore, that the X-ray spectra of VY\,Scl stars should be interpreted 
using the latter model, both in the high and low state (Mauche \&\ Mukai 2002).
This may have an impact on the suggested SSS connection (see also Patterson et al.\ 2001).

\subsection{AM CVn stars}
\label{amcvn}

The luminosities of the few AM\,CVn stars that have been detected 
in X-rays range from $\sim$10$^{28}$--5$\times$10$^{30}$\,erg\,s$^{-1}$
(Ulla 1995, and references therein; but see Verbunt et al.\ 1997).
The maximum of the overall flux distribution in AM\,CVn itself 
peaks around EUV wavelengths; there is no detectable hard X-ray emission 
(Ulla 1995). 
The X-ray luminosities of AM\,CVn stars agree with the coronal luminosities 
for single stars (e.g., Rosner et al.\ 1985; Hempelmann et al.\ 1995), 
and possibly with emission from single DB WDs 
(e.g., Fontaine et al.\ 1982). 

Of the few exceptions to the $F_{\rm X}$/$F_{\rm opt}$ versus 
$P_{\rm orb}$ relation (Sect.~\ref{overview}), is AM\,CVn.
With $P_{\rm orb}$$\sim$0.29\,h it has an 
unexpectedly small ratio of $\sim$0.002.
This small ratio might be explained by a high $\dot{M}$,
comparable to UX\,UMa stars (van Teeseling et al.\ 1996).
On the other hand, the AM\,CVn system GP\,Com has a ratio near unity
(van Teeseling \&\ Verbunt 1994). This may be due to the fact that the entire disk in GP\,Com
is in a (low) steady state, in which it will always be optically thin,
and will not undergo outbursts similar to that seen in DNe
(Marsh 1999).

V407\,Vul, a CV related to the AM\,CVn stars, 
was recently suggested to be a new type of double-degenerate CV
(Marsh \&\ Steeghs 2002). 
In this CV the mass transfer stream may hit
a non-magnetic WD directly due to a very compact orbit of 9.5\,min.
This results in pulsations in the X-ray flux every 9.5\,min,
with no X-ray emission in between pulses (suggestive for it being a polar
[see Sect.~\ref{polar}], 
however, neither polarization nor line emission is seen).
Its X-ray spectrum is soft 
($kT_{\rm bb}$$\simeq$40--55\,eV; Motch et al.\ 1996; 
Wu et al.\ 2002). 
This is explained by the stream breaking
into dense blobs which are able to penetrate the photosphere of the 
WD and therefore become thermalised, giving rise to the soft X-ray emission
(Marsh \&\ Steeghs 2002).

\section{X-ray emission from polars}
\label{polar}

\subsection{Introduction}

In polars, the originally
free-falling matter couples to magnetic field lines somewhere 
between the two stars and is guided to one or two accretion 
regions in the vicinity of the magnetic poles. These are
the sources of intensive X-ray radiation, mainly in the
soft X-ray regime, and of cyclotron radiation 
from IR to UV wavelengths. 
The observation of pulsed polarised radiation from the 
cyclotron source led to their
nick-names as polars (Krzemi\'nski \&\ Serkowski 1977). 

There is no recent review of the 
X-ray properties of polars in broad generality; the main
satellite-related aspects were reviewed 
({\it ROSAT}: Beuermann \& Thomas 1993; Beuermann \& Burwitz 1995; 
{\it EUVE}: Sirk \& Howell 1998; Mauche 1999; {\it ASCA}: Mukai 1995). Emission 
from post-shock flows in magnetic CVs is described by Cropper (1990).

Just two polars were known as variable stars before the era 
of X-ray astronomy began (AM\,Her, VV\,Pup) and a very 
few were detected in optical spectroscopic surveys before 1999
(AN\,UMa, CE\,Gru, MR\,Ser = PG1550+191). To date about 70 polars 
are known, the vast majority of them identified as counterparts of 
serendipitous X-ray sources. Only recently, much deeper optical
spectroscopic surveys ({\it Hamburg Schmidt telescope}, {\it SDSS})
have uncovered new systems in apparently permanent low states of accretion
(Reimers \&\ Hagen 1999; Reimers et al.\ 2000; Szkody et al.\ 2002b). 
They were not or just marginally detected in X-rays, and, due to their
low accretion rates, display intriguing cyclotron spectra. 

Polars are in the first place emitters of soft X-rays.
Therefore, the all-sky surveys conducted 
with {\it ROSAT} (XRT and WFC) in combination with optical identification 
programmes permitted for the first time a synoptic view 
of the CV sky with high sensitivity. 
Most polars are found below
the 2--3\,h CV period gap. The
gap itself is significantly filled in, possibly due to reduced braking
by trapping of the wind from the donor within the
magnetosphere of the WD (Webbink \& Wickramasinghe 2002).
The space density is of the order of $n_s$$\simeq$1--2$\times$10$^{-6}$\,pc$^{-3}$ 
of short-period systems and a
factor of 10 lower for long-period systems (Beuermann \& Schwope 1994).

\begin{figure*}[t]
  \vbox{\hskip -\leftskip
  \hspace{0.5cm}
  \psfig{figure=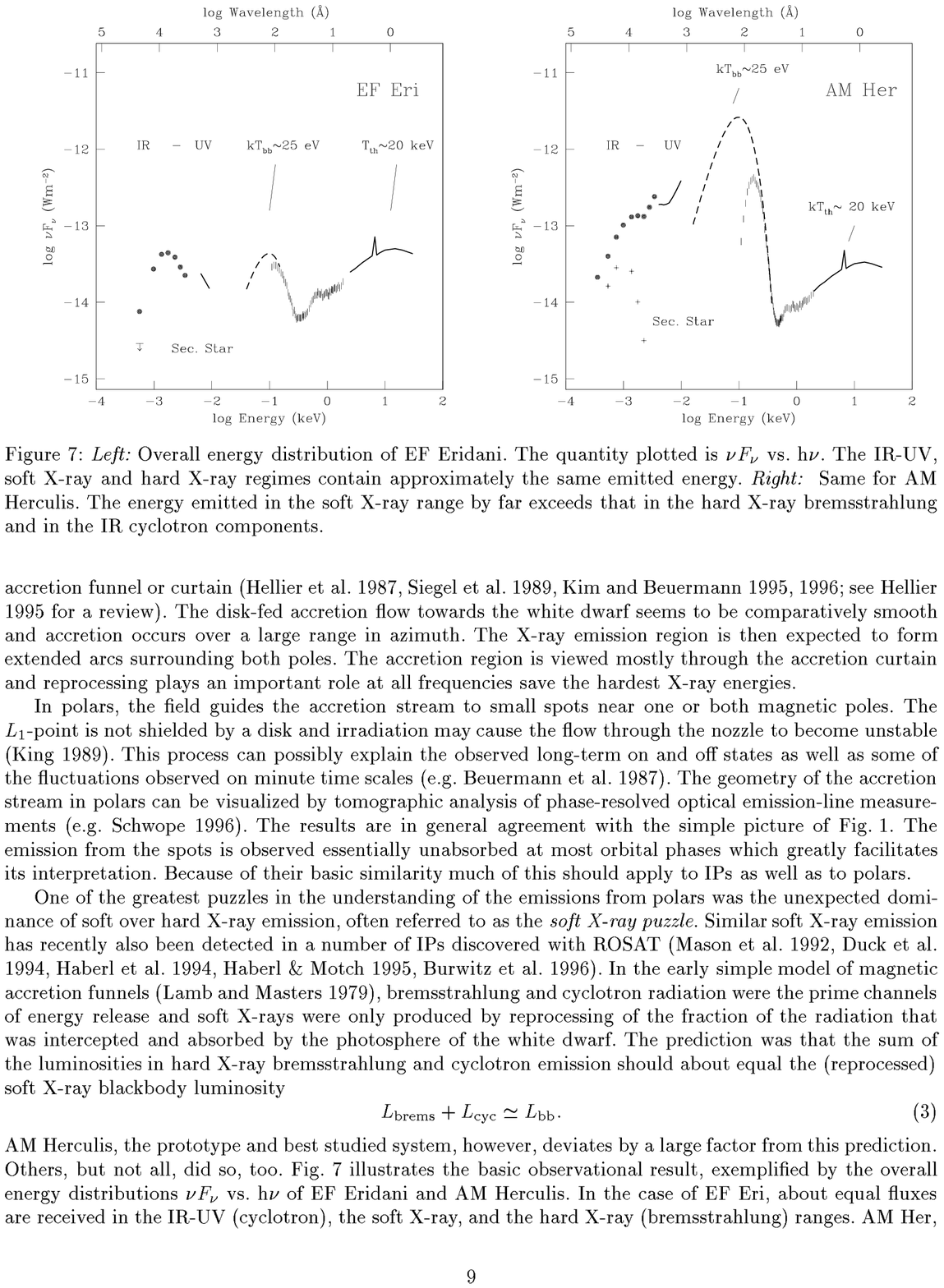,width=5.6cm,%
  bbllx=332pt,bblly=514pt,bburx=537pt,bbury=728pt,clip=}
  \psfig{figure=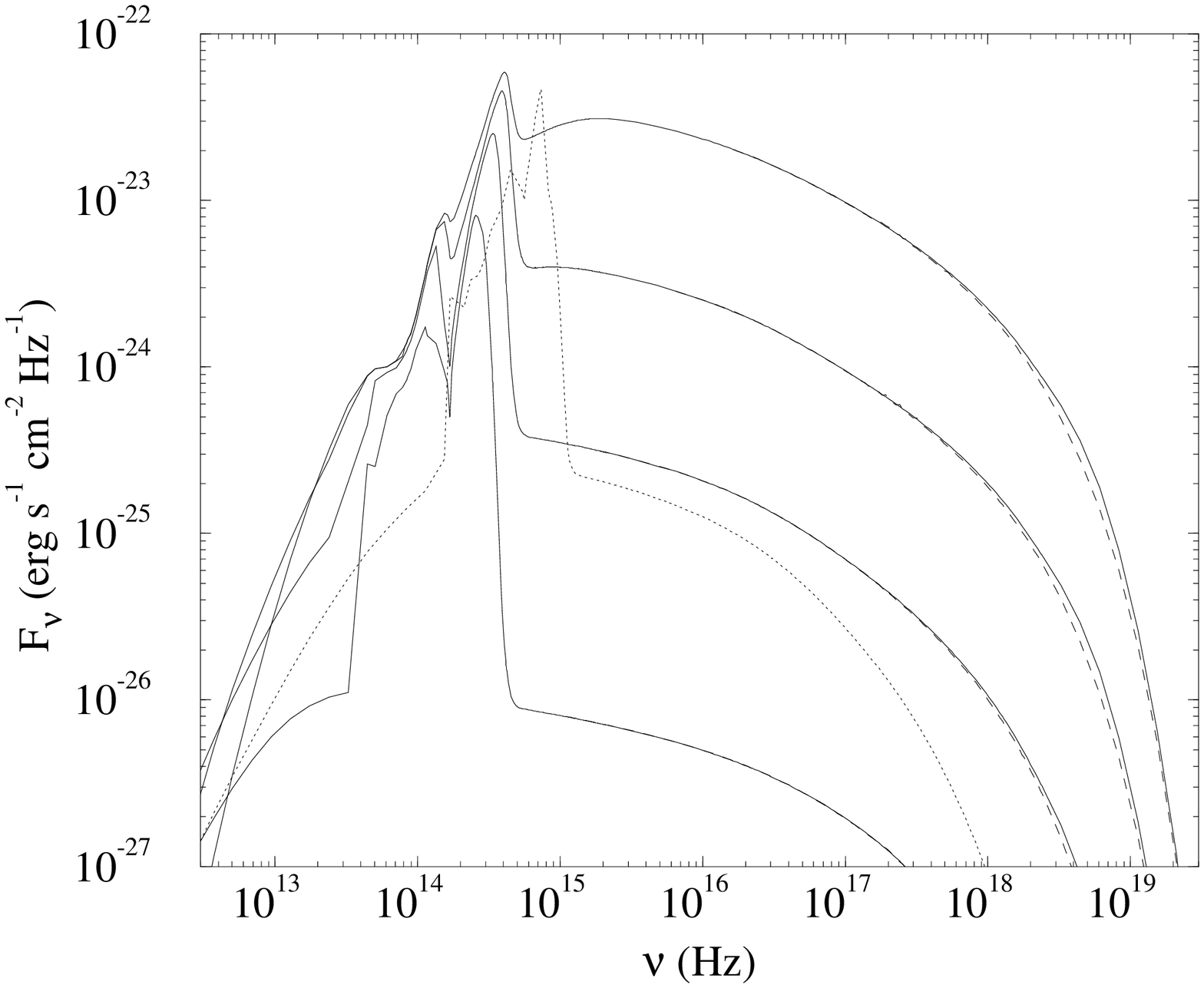,width=5.6cm,%
  clip=}}\par
  \caption{\label{f:amher_sed}
Observed (left, AM\,Her) and theoretical (right)
spectral energy distributions of polars (Beuermann 1999; Fischer \& Beuermann
2001). The models do not take into account the prominent soft X-ray component
of reprocessed origin, they account for the primary components
of optical cyclotron and hard X-ray bremsstrahlung radiation.
}
\end{figure*}

As shown in Fig.~\ref{f:amher_sed}, polars emit from the IR to the 
hard X-ray regime. Most of the radiation is accretion-induced.
A complete picture therefore requires multi-wavelength 
observations, preferably  obtained contemporaneously 
because of the inherent high variability on many time scales 
(from seconds to years).
X-ray observations are essential in order to determine 
the accretion scenarios (which 
requires a deconvolution of the X-ray spectra), the 
accretion geometries (which requires a deconvolution of the X-ray 
light curves), and the accretion history (which 
requires long-term monitoring in the X-ray domain). 

\subsection{Accretion-induced emission}

Matter in the accretion stream is accreted almost vertically 
on to the magnetic poles of the WD. The accretion process
is, therefore, almost always modeled in a one-dimensional quasi-radially
symmetric approximation. The presence of, e.g., accretion arcs with 
corresponding variety of accretion rates and deviations from the
radial symmetry due to inclined 
magnetic field lines is evident from observations but neglected in the
modeling for tractability.

Accretion is governed by three parameters, $M_{\rm WD}$, the accretion rate per unit area, 
$\dot{m}$, and $B$. The balance between those parameters determines, whether
the acretion region is heated via a strong hydrodynamic 
stand-off shock or by particle bombardment, and whether 
the cooling function is dominated by plasma emission 
or by cyclotron radiation. 

With appropriate boundary conditions, the equations of conservation
of mass, momentum and energy can be used to calculate the temperature and 
density as a function of height, as well as the emerging spectra. Only the 
one-dimensional non-magnetic case could be solved analytically (Aizu 1973).
Present numerical models include cooling by cyclotron emission by 
solving the fully frequency and angle-dependent radiative transfer 
and treat the accretion plasma in a 2-fluid approximation 
(Fischer \& Beuermann 2001; see Fig.~\ref{f:amher_sed}); they include gravity, 
account for its variation within the flow and use
up-to-date plasma emission codes (Cropper et al.\ 1998).
They take into account 
pre-shock heating and ionization, which influences the size 
of the shock jump and the formation of the emerging spectra.
However, current models are still one-dimensional and stationary, 
which limits their direct applicability to observational data.

Multi-temperature plasma emission models were fitted 
to the hard X-ray spectra of polars with the aim (among others) to estimate
$M_{\rm WD}$. A Compton reflection 
component from the irradiated WD surface, complex 
absorption (partial covering cold or warm absorbers) in the pre-shock 
flow or surrounding matter, fluorescent 
K$\alpha$ emission and cold interstellar absorption were taken into account
(e.g., Done \& Magdziarz 1998; Matt et al.\ 2000).
Such models give satisfactory fits to the data but tend to 
predict too high values of $M_{\rm WD}$ compared with 
dynamical mass estimates or parallaxes (Cropper et al.\ 2000). 
Van Teeseling et al.\ (1999) question the validity of coronal models
for the post-shock emission. High accretion-rate systems may be very
optically thick in the resonance lines, the corresponding 
asymmetric line emission may serve as a diagnostic tool
to probe the very inner accretion geometry.
X-ray line spectroscopy was used to infer sub-solar abundances
of the accreted matter (Done \& Magdziarz 1998; Ishida \& Ezuka 1999), in unresolved 
contrast to UV-line spectroscopy (Bonnet-Bidaud \& Mouchet 1987). 

In the `standard accretion model' (King \&\ Lasota 1979; Lamb \& Masters 1979) 
about half of the X-rays and of the cyclotron radiation 
are intercepted by the WD surface and are reprocessed 
as soft X-rays. This simple model predicts about equal luminosities
in the bremsstrahlung and cyclotron components on the one hand
and the soft emission on the other. Details of this balance depend 
on the hard X-ray albedo, the irradiation geometry and, observationally,
on the viewing geometry. However, since the early days ({\it EXOSAT} and {\it Einstein} 
era), a moderate to strong soft X-ray excess over the other components
was observed, creating what was referred to as the `soft X-ray puzzle'. 
The size of the 
soft excess was difficult to assess exactly but X-ray flux ratios 
$F_{\rm soft}/F_{\rm hard}$ up to $\sim$100 were reported. 
Difficulties to quantify the soft excess often 
arise from non-simultaneous observations in the soft and hard 
spectral bands and from incomplete spectral coverage 
of the soft component. At temperatures $kT_{\rm bb}$$\simeq$15--30\,eV
it has its peak emission in the EUV where most 
instruments have low sensitivity and interstellar absorption 
is severe. Apart from a few exceptions which indicate the 
presence of Ne\,{\sc VI} absorption edges or  Ne\,{\sc VII} and Ne\,{\sc VIII}
absorption lines, the soft spectra can be well described with a simple 
black body (Mauche 1999). The application of 
more physical models, e.g., pure-H or solar abundance stellar 
atmospheres, does not improve the fits due to the incompleteness
of the models and the low signal-to-noise of the data.

Two ways were proposed to cure the soft X-ray problem (Kuijpers \& Pringle 
1982). The first invokes shredding of the stream into diamagnetic 
blobs in the magnetospheric interaction region. Subsequent confinement 
and compression of the blobs leads to 
highly inhomogeneous accretion of filaments with partly or wholly buried 
shocks (Frank et al.\ 1988). The 
primary hard radiation cannot escape freely and the 
photosphere will be heated from below. Apart from 
hydrostatic computations of the temperature structure 
of a one-blob impact (Litchfield \& King 1990), this model
is not worked out in quantitative detail.

An alternative scenario applies to the low $\dot{m}$ and high $B$ case.
In such an environment 
cyclotron cooling becomes so efficient that it cools the 
plasma over a mean free path of the infalling particles, i.e., the 
shock is resolved and bremsstrahlung is suppressed (bombardment
solution, Woelk \& Beuermann 1996).
{\it ROSAT} observations of a large number of polars 
showed a clear relation between the size of the soft excess and the 
magnetic field strength in the accretion region
(Beuermann \& Schwope 1994; 
Ramsay et al.\ 1994)\footnote{Meanwhile, the field strength 
of about 45 systems have been measured. Some measurements
are based on Zeeman-split Balmer-lines from the photosphere or from 
an accretion halo (Schwope 1996), but most of these
measurements are based on the identification of cyclotron harmonic 
emission lines in low-resolution optical and/or IR spectra, originating 
from the accretion plasma at one or two accretion regions (for a review
see Wickramasinghe \& Ferrario 2000).}.
This correlation was explained either by enhanced fragmentation 
of the stream in the magnetosphere, i.e., by enhanced blobby accretion,
or by enhanced cyclotron cooling, thus supporting either of the 
two alternatives to the `standard' model. The decomposition of light 
curves in eclipsing systems (e.g., Bailey 1995) and the analysis of 
soft-to-hard X-ray cross-correlation functions (e.g., Beuermann et al.\ 1991) 
suggest that regions with low and high $\dot{m}$ co-exist. Consequently, 
the accretion region cannot be described in terms of just one of the 
scenarios. Further modification to the `standard' model 
arises from the fact, that in AM\,Her the reprocessed component 
is observed with the right energy content and is observed in the 
UV instead of the soft X-ray regime, suggesting that the soft 
X-rays are completely decoupled from the other radiation processes 
(G\"ansicke et al.\ 1995).
Similar multi-band investigations which include particularly the 
UV spectral regime are missing for other polars.

\subsection{X-ray light curves}

\begin{figure}
\hspace{-1.0cm}
\resizebox{0.5\hsize}{!}
{\includegraphics[bbllx=28pt,bblly=250pt,bburx=552pt,bbury=772pt,clip=]{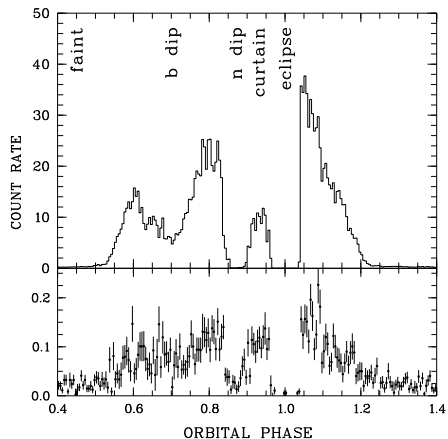}}
\hfill
\parbox[b]{0.52\hsize}{
\caption{\label{f:hulc} 
{\it ROSAT} X-ray light curves of the high-inclication polar HU\,Aqr.
The original photon data comprise 36\,ksec and were phase-averaged over the
125\,min orbital period. Phase zero corresponds to superior
conjunction of the WD. The upper panel encompasses the whole 
spectral band pass, the lower panel only the hard X-ray spectral 
component (0.5--2.0\,keV). Several important features around
the orbital cycle are indicated. Adapted from Schwope et al.\ (2001).
\vspace{0.3cm}
}
}
\end{figure}

Polars display a rich phenomenology of X-ray light curves
despite their rather simple accretion geometry. The light curves
offer large  diagnostic potential, since they 
are modulated by the location and the three-dimensional 
extent of one or several accretion spots,
by stellar eclipses (11 out of 70 systems display stellar eclipses), 
by nonstationary accretion processes, and
by absorption of X-rays within the binary.
The long-term light curves are affected by shifts of the accretion regions
in longitudinal and/or lateral directions, by 
changes between one- and two-pole accretion modes (which gives those 
systems a completely different appearance) and by large-scale
variations of $\dot{M}$. These long-time
dependencies occur at unpredictable moments and for unpredictable 
duration. It is a common assumption that long-term changes of the
accretion rate are related to star spots 
on the donor at the $L_1$ (King \& Cannizzo 1998) and an attempt has 
been made to reconstruct a possible pattern of star spots from the 
accretion history of AM\,Her (Hessman et al.\ 2000). While 
irradiation-induced structure is obvious in Doppler tomograms 
of the donors, these experiments have failed so far to make 
star spots visible (Schwope 2001). 

The main features of the X-ray light curve of a polar in a one-pole 
accretion geometry are exemplified in Fig.~\ref{f:hulc}, the {\it ROSAT} 
light curve of HU\,Aqr (Schwope et al.\ 2001). 
While interpreting these 
light curves one should bear in mind
that the spin of the WD is synchronously locked with $P_{\rm orb}$, 
typically to a degree better than $10^{-6}$.
The light curve displays several pronounced
features, labeled as broad dip (`b dip'), narrow
dip (`n dip') and eclipse.
The light curve also 
shows a pronounced on-off pattern due to self-occultation 
of the accretion region behind the WD. In the faint 
phase some residual X-ray emission is present, which is probably of scattering 
origin. The length and 
phasing of the bright phase allows clues to be drawn on the 
latitude and longitude of the accretion spot. 
Most polars have the main accretion spot in the quadrant of the 
ballistic stream (seen by a hypothetical observer on the  
WD), i.e., on the leading side of the WD. 
However, a few systems have their spots on the trailing 
side of the WD or even on the opposite from the donor.
This implies complex motions of matter in the magnetosphere 
and it is by no means clear to what extent the simple picture of accretion
via Roche-lobe overflow at $L_1$ is applicable. Recent 
Doppler tomograms show that even the putative ballistic, freely falling
stream in the vicinity of the $L_1$ might be influenced by the 
magnetic fields in the binary (Schwarz et al.\ 2002). 
The spots do not show any preferred latitude, i.e., the spin axis
of the WD seems not to be aligned with the rotation axis. 

While the hard X-rays are assumed to be formed
by (mainly) optically thin plasma radiation the X-ray brightness
should be constant throughout the bright phase. Instead, large fluctuations 
are seen which are assigned to nonstationary accretion and occultations.
The soft X-rays originate from optically thick surfaces
and orbital modulations are expected (and observed) to be more marked.
The bright phase in general is far from being 
compatible with a flat accretion spot and heating by accretion 
blobs rather than irradiation plays an important role.
The depth-dependent temperature structure
in the vicinity of an accretion filament was computed 
by Litchfield \& King (1990) using similarity to a heat
conduction problem. The photosphere is then assumed to be shaped
like a mound, but its height proved not to be sufficient to 
reproduce soft X-ray light curves of, e.g., the 
anomalous state of AM\,Her (Heise et al.\ 1985), 
suggesting that hydrodynamic 
splashes rather than a hydrostatic atmosphere 
are more likely responsible for the soft X-ray light curves.

The bright phase may undergo shifts with respect 
to binary phase zero. In some cases this 
reflects the accretion rate dependent penetration of the 
magnetosphere by the ballistic stream before it gets threaded onto magnetic
field lines. In other cases it indicates a small asynchronism 
of $P_{\rm spin}$ and $P_{\rm orb}$ (Schwope et al.\ 2001, 2002).

There are 11 eclipsing systems; two of them, UZ\,For and HU\,Aqr,
were found bright enough 
to resolve the ingress into or the egress from the eclipse in 
soft X-rays. Detailed modeling
shows, that the soft X-ray accretion spots have an angular lateral 
extent of less than $5^{\circ}$ (Warren et al.\ 1995; Schwope et al.\ 2001)
and a vertical extent of less than 0.05\,R$_{\rm WD}$.
Hence the soft X-ray accretion region is pillbox-shaped rather than 
pencil-shaped.

The narrow dip is due to absorption in 
the transient accretion stream, which 
becomes X-rayed when it has just left the orbital 
plane. It can be observed in systems in which the orbital 
inclination exceeds the co-latitude of the accretion spot.
Its phase indicates the azimuth and its 
width the size of the threading region in the magnetosphere.
To a first order, the coupling region is located where 
the magnetic pressure overcomes the ram pressure 
of the ballistic stream. A simultaneous hard X-ray and near-IR study
of the dips in EF\,Eri (Watson et al.\ 1989) demonstrates the 
presence of substantial structure in the dips, implying significant
density fluctuations in the stream, either spatial or temporal.
The origin of the broad dip centred at an earlier phase 
remains unclear so far. Its width and X-ray 
colour suggest an origin in warm absorbing matter in close vicinity
to the hot accretion spot. Its diagnostic potential needs to be explored.

\section{X-ray emission from intermediate polars \label{IPs}}
\label{IP}

\subsection{Introduction}
\label{intro_IP}

The first of the asynchronous magnetic CVs to be discovered was the remnant of
Nova Herculis 1934 -- DQ\,Her -- in 1954. Walker (1956) found a highly stable
71.1\,s optical modulation with amplitude of a few hundredths of a magnitude,
which disappeared during eclipse. In analogy with the models for X-ray pulsars
(Pringle \& Rees 1972; see Chapter 7) an accreting oblique dipole rotator was eventually
suggested (Bath et al.\ 1974a). The correctness of this model was
demonstrated by the discovery of a phase shift in the 71\,s signal during
eclipse that could be matched to the eclipse of a beam of high energy
radiation, emitted by the rotating WD, as it swept over the accretion disk
and was reprocessed into optical wavelengths (Warner et al.\ 1972; 
Patterson et al.\ 1978). 

The evident success of this accreting magnetic model led to its adoption 
when the first X-ray CVs to have two simultaneous periodic 
modulations were found. For example, AO\,Psc was observed to have a 
strong 14.3\,min optical periodicity (Warner 1980; Patterson \&\ Price 1980) 
but a 13.4\,min modulation in hard X-rays (White \& Marshall 1980). The 
realisation that the frequency difference of these two modulations is equal to 
the orbital frequency showed that the X-ray period arises from rotation of 
the WD, but the optical modulation must be caused by the rotating beam 
being reprocessed from some structure fixed in the rotating frame of the 
binary (e.g., the donor or the thickening of the disk where the stream 
impacts). It was therefore recognised that only two `clocks' are really 
present, i.e., $P_{\rm orb}$ and the spin period of the WD, $P_{\rm spin}$. 

Because of amplitude modulation, caused largely by geometrical projection
effects, a suite of modulations is sometimes seen in optical observations. 
Denoting $\Omega$=2$\pi/P_{\rm orb}$ and $\omega$=2$\pi/P_{\rm spin}$, the
frequencies $\omega$$-$$\Omega$, $\omega$$-$2$\Omega$, $\omega$+$\Omega$ and
$\omega+2\Omega$ are predicted to occur (Warner 1986). The importance of these
sidebands is that they act as proxies for X-rays in those cases where no X-ray
modulation has been observed; there are no models other than IPs that explain
the presence of orbital sidebands.

\begin{figure*}[t]
\setlength{\unitlength}{1cm}
\hspace{1cm}
\begin{picture}(10,7)
\put(0,7.5){\includegraphics{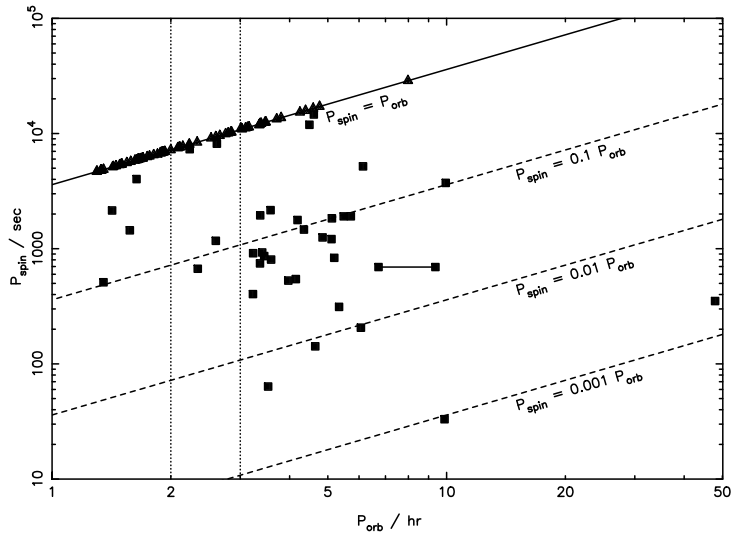}}
\end{picture}
 \caption{\label{pp}
   The $P_{\rm spin}$ versus $P_{\rm orb}$ diagram of magnetic CVs. Polars are indicated
   by triangles; IPs are indicated by squares.}
\end{figure*}

Direct observation of an X-ray modulation, usually denoting $P_{\rm spin}$,
accompanied by $P_{\rm orb}$, usually obtained from optical photometric or
spectroscopic observations, is required to give full conviction to
classification as an IP. But an optical periodicity of proven stability (to
distinguish from the quasi-periodic oscillations discussed in Sect.~\ref{oscillations}), and the presence of one
or more orbital sidebands, even without any X-ray detection at all, give an
irresistible urge for inclusion in the IP lists.  
The parameter space occupied by the IPs is illustrated in Fig.~\ref{pp}.
    
\subsection{Modes of accretion}

Most IPs are expected to accrete via some form of truncated accretion disk
whose inner edge is at the magnetospheric radius. From here, material will
attach onto field lines and flow towards the magnetic poles, forming `accretion
curtains' above each pole (Rosen et al.\ 1988).  Unlike the
polars, the accretion flow impacting the WD in IPs will therefore be more
extended, occurring over a greater fraction of the WD surface. At some point in
the flow, the material will experience a strong shock before settling and
cooling mainly by thermal bremsstrahlung. This region is thus the origin of the
observed X-ray emission. Modulation at the WD spin period is produced by a
combination of self occultation and varying photo-electric absorption towards
the X-ray emission sites.

Some IPs are also believed to accrete (at least in part) directly via a stream,
in a similar manner to polars. This stream may overflow a disk (disk-overflow
accretion; e.g., Hellier et al.\ 1989; King \& Lasota 1991; Armitage \& Livio 1996) or
replace a disk entirely (stream-fed accretion; Hameury et al.\ 1986). 
As the stream flips from pole to pole, this will naturally give rise to an
X-ray modulation at the sideband frequency $\omega$$-$$\Omega$ (Hellier 1991;
Wynn \& King 1992; Norton 1993).  In disk-fed accretion it is likely that the
footprints of the field lines onto which the flow attaches are semi-circular
X-ray emitting arcs around each magnetic pole, and are fixed on the surface of
the WD. By contrast, the field lines to which the stream-fed or
disk-overflow accretion attaches are likely to have smaller footprints at each
magnetic pole and these will `migrate' around the pole to follow the incoming
stream, as a function of the sideband phase (Norton et al.\ 1997). 

\subsection{X-ray lightcurves}

In early X-ray observations, the light curves of IPs folded at the WD spin
period were seen to be roughly sinusoidal and interpreted as largely due to
self occultation of the emission area by the WD (King \& Shaviv 1984). First
hints that this was not the case came with {\em EXOSAT} observations (Mason
1985; Watson 1986; Norton \& Watson 1989) which showed that the modulation
depth tended to increase with increasing X-ray energy over the range 
1--10\,keV, indicating that photo-electric absorption made some contribution to the
observed modulation.

The now widely accepted `accretion curtain' model was proposed to
explain the data from EX\,Hya (Rosen et al.\ 1988, 1991).
In this model, the
emission region is a tall, thin, arc-shaped curtain and the largest X-ray flux
is seen when the curtains are viewed from the side (i.e., when a given pole is
pointing {\em away} from the observer). Although EX\,Hya is an atypical IP, such
a model was also successfully applied to the other IPs (e.g., Hellier et al.\ 1991).  

\begin{figure*}[t]
\setlength{\unitlength}{1cm}
\hspace{0.65cm}
\begin{picture}(10,7)
\put(0,0){\includegraphics{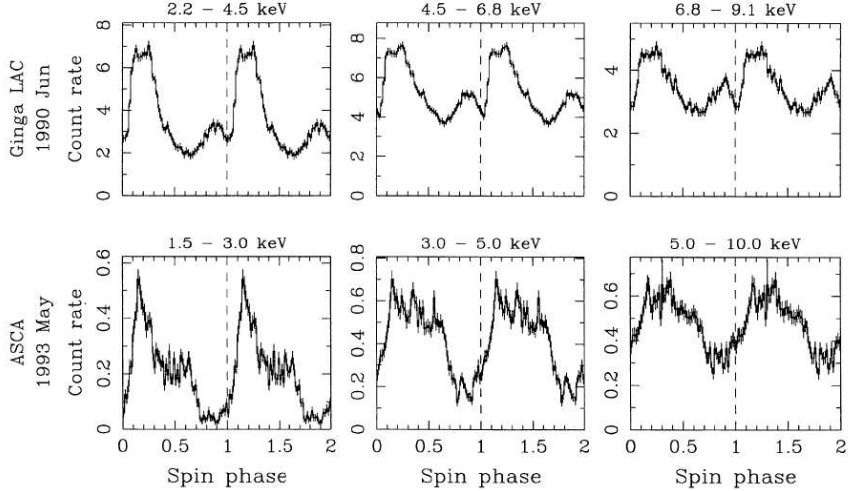}}
\end{picture}
  \caption{\label{pulse}
   X-ray pulse profiles of FO\,Aqr. Adapted from Beardmore et al.\ (1998).}
\end{figure*}

{\em Ginga}, {\em ROSAT}, {\em ASCA} and {\em RXTE} each observed many IPs,
producing lightcurves with extremely high signal-to-noise in many cases. Whilst
the pulse profiles of some sources (e.g., EX\,Hya; Rosen et al.\ 1991) still
appear roughly sinusoidal, there are indications that additional structure may
be present in others. For example, the combined {\em ROSAT} and {\em Ginga}
pulse profiles of AO\,Psc and V1223\,Sgr show evidence for a small notch
superimposed on the peak of the pulse (Taylor et al.\ 1997).  Several objects,
including GK\,Per (Ishida et al.\ 1992), XY\,Ari (Kamata \& Koyama 1993), V405\,Aur
(Allan et al.\ 1996), YY\,Dra and V709\,Cas (Norton et al.\ 1999) show pulse
profiles that are double peaked (at least on some occasions they are observed).
Some of the most complex pulse profiles are those seen from FO\,Aqr (Norton et
al.\ 1992a; see Fig.~\ref{pulse}), BG\,CMi (Norton et al.\ 1992b) and 
PQ\,Gem (Duck et al.\ 1994) which show narrow notches superimposed on broader modulations and
pulse profiles that change significantly with $\phi_{\rm orb}$.  Many of these
X-ray pulse profiles also vary dramatically on time scales of months or years.
For instance, those of GK\,Per and XY\,Ari are single peaked in outburst 
(Watson et al.\ 1985; Hellier et al.\ 1997) but double peaked in
quiescence; in other cases, such as V709\,Cas, the contributions of various
harmonics of the spin frequency are seen to vary (de Martino et al.\ 2001).

It has been noted (e.g., Norton et al.\ 1999) that the IPs exhibiting
double-peaked X-ray pulse profiles are mostly those with short $P_{\rm spin}$. 
The WDs in these objects therefore probably have weak magnetic fields, so the
magnetospheric radius is relatively small. Consequently the footprints of the
disk-fed accretion curtains on the WD surface are relatively large. In contrast
to a conventional accretion curtain, the optical depths to X-ray emission are
therefore lowest in the direction along the magnetic field lines, and highest
in the direction parallel to the WD surface, such that the emission from the
two poles conspires to produce double-peaked X-ray pulse profiles (Allan et al.\
1996; Hellier 1996; Norton et al.\ 1999). Such a pulse profile is therefore {\em
not} a unique indicator of two-pole accretion. Indeed, two-pole
accretion onto smaller regions of the WD surface may be considered the `normal'
mode of behaviour in a disk-fed IP with a longer $P_{\rm spin}$ (and
therefore a higher field strength), resulting in a single-peaked pulse profile.
Indications of the size of the X-ray emitting region in IPs have come from a
study of the deeply eclipsing IP XY\,Ari. The $\lesssim$2\,s egress from eclipse
seen in {\em RXTE} data limits the accretion region to $\lesssim$0.002 of the WD
surface area (Hellier 1997).

Whereas fast rotators with relatively weak fields show double-peaked pulse
profiles, several slower rotators with larger fields (and therefore larger
magnetospheres) have been seen to exhibit an X-ray sideband modulation (i.e., at
a frequency $\omega$$-$$\Omega$) at some time. 
A strong sideband signal is seen in TX\,Col (Buckley \& Tuohy 1989),
and FO\,Aqr (Norton et al.\ 1992a), and weaker signals
in AO\,Psc and V1223\,Sgr (Hellier 1992). A dominant sideband
period may exist in BG\,CMi (Norton et al.\ 1992b).  
Observations of V2400\,Oph subsequently confirmed this system as the first truly
diskless IP by revealing that its X-ray signal varies {\em only} at the 1003\,s
sideband period (Buckley et al.\ 1997; Hellier \& Beardmore 2002).

The relative strengths of the X-ray sideband and spin modulation in FO\,Aqr have
been seen to vary on time scales of years (Beardmore et al.\ 1998) 
and those in TX\,Col on time scales as
short as months (Norton et al.\ 1997). 
The interpretation is that the relative amounts of accretion occurring via a
stream and via a disk vary, possibly due to changes in $\dot{M}$ or other
activity on the donor near to the $L_1$ point. 

IPs sometimes show strong orbital modulations in their X-ray light curves. The
compilation by Hellier et al.\ (1993) showed
that FO\,Aqr, EX\,Hya, BG\,CMi and AO\,Psc all have orbital dips characterised by
increased photo-electric absorption around $\phi_{\rm orb}$$\sim$0.8. Such orbital
modulations have been confirmed by subsequent observations 
(e.g., Norton et al.\ 1992a,b; Taylor et al.\ 1997; Allan et al.\ 1998). 
Hellier et al.\ (1993) concluded that
the cause was likely to be similar to that in DNe and LMXBs and due
to material thrown out of the orbital plane by the stream impact with the disk
or the magnetosphere (see Sect.~\ref{quiescence}). 
Alternatively, or additionally, a spin pulse profile
that varies with $\phi_{\rm orb}$ (such as will arise naturally in a stream-fed or
disk-overflow model) will naturally give rise to an orbital modulation. It is
likely that this effect contributes to the observed orbital modulation, at
least in some systems.

\subsection{X-ray spectra}

The relatively low signal-to-noise and low resolution X-ray spectra of IPs
obtained with {\em EXOSAT} were adequately fitted with single temperature 
($\sim$10's of keV) bremsstrahlung continua passing through a partial absorber which
varied with phase (see the compilation by Norton \& Watson 1989). Fluorescent
Fe\,K$\alpha$ lines were also seen in most IPs (Norton et al.\ 1991). 
{\em Ginga} observations largely confirmed these results but showed that both
thermal and fluorescent contributions to the Fe lines were present (Ishida
1991). The higher spectral resolution of {\em ASCA} allowed more emission lines
to be detected in the X-ray spectra of EX\,Hya (Ishida \& Fujimoto 1995) and 
AO\,Psc (Fujimoto \& Ishida 1995; Hellier et al.\ 1996), 
for example, and also showed
that V405\,Aur, PQ\,Gem, AO\,Psc, BG\,CMi, V2400\,Oph, TV\,Col and V1025\,Cen have 
thermal Fe\,K$\alpha$ lines that are broadened by $\sim$200\,eV 
(Hellier et al.\ 1998). In each case, up to three Gaussian lines were required, 
corresponding to cold, H-like and He-like Fe.

A significant advance in the modelling of IP X-ray spectra came with models
that used a multi-temperature emission region, including effects such as
reflection from the surface of the WD and partially ionised absorbers 
(e.g., Cropper et al.\ 1998; Beardmore et al.\ 2000). Using this
technique, Cropper et al.\ (1998) fitted the {\em Ginga} spectra of 9 IPs and
determined WD masses for them.

{\em ROSAT} discovered a sub-class of IPs, e.g., V405\,Aur, PQ\,Gem and UU\,Col,
characterised by soft X-ray spectra, with black-body components at 
$T_{\rm bb}$ of $\sim$10's of eV (Mason et al.\ 1992; Haberl et al.\ 1994; Haberl \& Motch 1995;
Burwitz et al.\ 1996). The soft X-rays from these objects probably
originate, as in polars, from the heated WD surface around the accreting poles.
The black-body fluxes indicate fractional areas of only $\sim$10$^{-5}$ of the
WD surface for the soft X-ray emission region (Haberl \& Motch 1995).

At the time of writing, {\em Chandra} and {\em XMM-Newton} spectra of IPs are
just becoming available and will probably revolutionize our understanding of IP
X-ray spectra. For example, Mauche (2002b) shows that 
line ratios from the {\em Chandra} spectrum of EX\,Hya may be used to
determine a plasma temperature which spans the range 0.5 to 10\,keV and a
plasma density $n$$\gtrsim$2$\times$10$^{14}$\,cm$^{-3}$. Mukai et al.\ (2003)
demonstrate that EX\,Hya's {\em Chandra} spectrum is well fit by a simple
cooling-flow model, as are those of the DNe U\,Gem and SS\,Cyg, and the old nova V603\,Aql.
In contrast, the {\it Chandra} spectra of the IPs V1223\,Sgr, AO\,Psc and 
GK\,Per are inconsistent with such a model, but conform with the expectations
for line emission from a photo-ionised plasma.

\section{Rapid oscillations}
\label{oscillations}

\subsection{Dwarf nova oscillations}
\label{dno}

\subsubsection{Introduction}

The rapid oscillations seen in DQ\,Her (see Sect.~\ref{intro_IP}) and AE\,Aqr are of very high
stability ($Q$=$P$/$\dot{P}$$>$10$^{12}$). In contrast, optical oscillations of low
stability ($Q$$\sim$10$^{4-6}$) were discovered in some high $\dot{M}$ CVs (namely, DNe
in outburst and NLs) by Warner \&\ Robinson (1972). These are
known as dwarf nova oscillations (DNOs) and are usually of very low
amplitude (typically less than 0.01\,mag) and span the range 5--100\,s,
with a concentration near 25\,s. In a given CV they always appear at
similar periods. Their short periods indicate a source near to the
WD, and there was early expectation that they would be found at short
wavelengths. They were indeed later found in the soft X-ray, EUV and UV
regions. Only recently, however, have simultaneous EUV and optical
observations of DNOs shown that, again as expected, the same phenomenon is
being observed in all wavelength regions (Mauche \&\ Robinson 2001) -- but
frequently the optical modulated flux is merely reprocessed X-ray and EUV
radiation.

Before discussing observations of X-ray DNOs we describe a physical model
that is gaining acceptance as an explanation of the DNOs. In its essence
it is an IP model, but the magnetic field lines are
connected to the accreted material near the equator of the WD, and
not rooted in its interior. Paczy\'nski (1978) pointed out that if the
intrinsic field of the WD is low enough ($B$$\lesssim$10$^{5}$\,G: Katz 1975) the
accreted material will be able to circulate around the equator of the
WD. (The high $Q$ of DQ\,Her shows that in that CV the field is strong
enough to lock the exterior layers to the interior, so the accretion
torque is applied to the entire WD.) The shear in the accreting
equatorial belt may generate a field strong enough to control the gas flow
near the surface of the WD -- but $P_{\rm spin}$ of the belt is
determined by magnetic coupling to the inner edge of the disk. As $\dot{M}$
waxes and wanes during a DN outburst the inner radius (and
Keplerian period) of the disk is first reduced and then increased. The
result is a low inertia magnetic accretor (Warner \&\ Woudt 2002), which
explains the large range of a DNO period during a DN outburst, and
why it is observed to reach a minimum value at the maximum of $\dot{M}$.
There is direct spectroscopic evidence for rapidly spinning equatorial
belts in DNe during outburst (e.g., Sion et al.\ 1996).

\subsubsection{Soft X-ray DNOs}

Our knowledge of the oscillations in soft X-rays (0.1--0.5\,keV) during
outbursts of DNe is dominated by observations made of SS\,Cyg, 
U\,Gem and VW\,Hyi; all relatively nearby and optically bright objects. The
first detections were in SS\,Cyg (C\'ordova et al.\ 1980b, 1984), U\,Gem 
(C\'ordova et al.\ 1984) and VW\,Hyi (van der Woerd et al.\ 1987). Modulated soft X-ray
emission appears in all of the observed SS\,Cyg outbursts, but in only one
of three observed U\,Gem outbursts. The hard X-ray emission in SS\,Cyg is
not modulated (Swank 1979). The soft X-ray modulation amplitudes are much
greater than in the optical; generally $\sim$25\%, but as much as 100\%\ for
individual cycles. This shows that much of the accretion luminosity is
involved in the modulation process. Other DNOs have
been detected, i.e., in the DN HT\,Cas and the NLs YZ\,Cnc, RW\,Sex
and AB\,Dra (C\'ordova and Mason 1984), but the last three are more probably
of the quasi-periodic type discussed in Sect.~\ref{QPO}.
Table~\ref{xrdno} lists the published studies of the three bright DNe.

\begin{table}
\caption{EUV and Soft X-ray DNOs in DNe during outburst.}
\begin{tabular}{cccc}
\hline
Star & $P_{\rm orb}$\,(h) & Period (s) & References \\
\hline
SS\,Cyg  & 6.60 & 9 & C\'ordova et al.\ (1980b) \\
 & & 10.7 & C\'ordova et al.\ (1984) \\
 & & 9.6--10.1 &  Watson et al.\ (1985)\\
 & & 7.4--10.4 & Jones \&\ Watson (1992) \\
 & & 7.2--9.3 & Mauche (1996a) \\
 & & 2.8$^{\ast}$ & van Teeseling (1997b) \\
 & & 2.9--8.2 & Mauche \&\ Robinson (2001) \\
 & & 9.1 & Mauche (2002a) \\
U\,Gem & 4.25 & 25--29 & C\'ordova et al.\ (1984) \\
 & & $\sim$25 & Long et al.\ (1996) \\
VW\,Hyi & 1.78  & 14.06--14.4 & van der Woerd et al.\ (1987) \\
\hline
\multicolumn{4}{l}{\footnotesize $^{\ast}$A frequency doubling had occurred.} \\
\end{tabular}
\label{xrdno}
\end{table}

The SS\,Cyg observations by C\'ordova et al.\ (1980b, 1984) provided the first
means of analysing the short-term temporal variations of the DNOs; in the
X-ray region individual cycles can be seen, whereas in optical
observations the DNOs are only seen in Fourier transforms (FTs). However, later
observations of rare large amplitude optical DNOs, especially those in the
DN TY\,PsA (Warner et al.\ 1989), showed behaviour
similar to that in X-rays, namely that the DNOs maintain relatively high
coherence for a time $\Delta T$ and then jump suddenly ($t$$<$100\,s) to a period
typically 0.02\,s different. These jumps in period can be in either
direction and are superimposed on the steady increase or decrease in
period associated with, respectively, decreasing or increasing luminosity.
When the luminosity is not changing rapidly, i.e., in NLs or in
DNe near maximum light, $\Delta T$ can be in excess of an hour; but late
in an outburst $\Delta T$ decreases to hundreds of seconds and the DNOs become
incoherent and difficult or impossible to detect with FT
techniques.

In the past five years considerable progress has been made in two areas:
the observation of DNOs in the EUV flux (which may be assumed to be a proxy
for X-ray modulation) and the extension of studies in the optical. Mauche
(1996a, 1997) and Mauche \&\ Robinson (2001) have studied the EUV during
outbursts of SS\,Cyg and discovered several new phenomena, including a
frequency doubling of DNOs near maximum of outburst. This shows as a
reduction of DNO period from $\sim$6\,s to $\sim$3\,s; X-ray observations made near
maximum of a different outburst of SS\,Cyg also showed the $\sim$3\,s modulation
(van Teeseling 1997b). It is possible that the effect is in essence
geometrical, with emission from two accretion poles being seen when the
inner edge of the disk is very close to the WD surface (Warner \&\ Woudt 2002).

Optical studies of VW\,Hyi (Woudt \&\ Warner 2002; Warner \&\ Woudt 2002) show
the correlation of DNO period with luminosity, including for the first
time detecting oscillations at $\sim$14\,s near maximum, which were previously
only seen in X-rays (van der Woerd et al.\ 1987). A rapid slowdown of the
DNOs, from a period $\sim$20\,s to $\sim$40\,s over about 6\,h, coincides with the
epoch when the EUV flux plummets almost to zero, and is interpreted as a
propeller phase in which accretion is prevented by the magnetic field
attached to the rapidly spinning equatorial belt. Following the propeller
phase a frequency doubling occurs, which may be a change (at least in
visibility) from single pole to two-pole accretion.

X-ray modulations at 27.87\,s in WZ\,Sge in quiescence (Patterson et al.\ 1998) 
supports the magnetic accretion model for their origin (see, e.g.,
Warner \&\ Woudt 2002, and references therein). Their behaviour in the UV during superoutburst 
(Knigge et al.\ 2002), for example the lack of coherence and the
occurrence of harmonics, also resembles DNO behaviour.

\subsection{Quasi-Periodic Oscillations}
\label{QPO}

In 1977 a second class of unstable optical oscillations was found during
dwarf nova outbursts; these have $Q$$\sim$5, which means that (as they are
spread over a wide range of frequency) they are hard to detect
in FTs and were only noticed because of large amplitude in the light curve
(Patterson et al.\ 1977). Their low coherence gives them the
name `quasi-periodic oscillations' (QPOs). They have time scales typically
an order of magnitude longer than the DNOs, can be present or absent
during outbursts, and are independent of whether DNOs are active. They
have been commonly seen in DN outbursts and in NLs, and even occasionally in DNe at quiescence (Warner
1995; Woudt \&\ Warner 2002).

Very few observations of CV QPOs in X-rays have been made. In
soft X-rays a very low amplitude signal at 83\,s in 
SS\,Cyg during one outburst was found, and 111\,s in another (Mauche 1997, 2002a); 
C\'ordova \&\ Mason (1984) found a 12\%\ amplitude modulation at 585\,s in U\,Gem during
outburst and Ramsay et al.\ (2001a) found a modulation at 2240\,s at low
energies in OY\,Car just after the end of an outburst. At higher energies
(2--10\,keV) large amplitude $\sim$500\,s
modulations in VW\,Hyi in the final stages of decline from an outburst
were found (Wheatley et al.\ 1996b).

A third class of optical oscillations has recently been recognised
(Warner et al.\ 2003), which are present during outburst and
even very occasionally in quiescence, and change very little in period
throughout the outburst of a DN. They are represented by the 
$\sim$88\,s oscillations in VW\,Hyi (Haefner et al.\ 1979), the $\sim$33\,s
oscillations in SS\,Cyg (Patterson 1981) and the $\sim$92\,s oscillations in the
NL EC\,2117$-$54 (Warner et al.\ 2003). They are
typically a factor $\sim$4 longer in period than the standard DNOs. 
They may be represented in X-rays by the $\sim$130\,s
periodicities seen in U\,Gem during outburst (C\'ordova et al.\ 1984; C\'ordova \&\ Mason 1984).

From the rich phenomenology of optical DNOs and QPOs in dwarf nova
outbursts (especially VW\,Hyi) it has been suggested that the QPOs are
caused by slow progradely travelling waves in the inner disk; probably
close to the inner radius where magnetic channelling begins 
(Warner \&\ Woudt 2002). Then the optical QPOs are caused by `reflection' and
obscuration of radiation from the central regions of the disk and WD.

In the X-ray region quasi-periodic obscuration would account for the
modulations listed above. In all of those cases the X-ray QPOs are similar
in time scale to the optical QPOs in the same stars; no simultaneous
observations in optical and X-rays have yet been made. The newly
recognised longer period DNOs probably arise from magnetically controlled
accretion onto the body of the WD itself (rather than just its
equatorial belt).

\begin{figure}[t]
\hspace{-1.2cm}
\resizebox{0.52\hsize}{!}
{\includegraphics[bbllx=17pt,bblly=175pt,bburx=548pt,bbury=674pt,clip=]{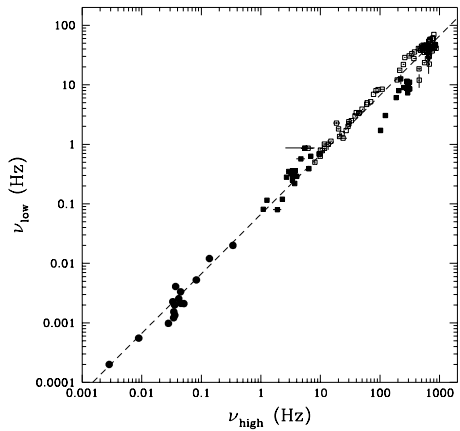}}
\hfill
\parbox[b]{0.5\hsize}{
\caption{\label{qpodno}
The `two-QPO diagram' for LMXBs (filled squares: black 
hole binaries; open squares: neutron star binaries)
and 17 CVs (filled circles). The LMXB data are from 
Belloni et al.\ (2002). The dashed line marks $P_{\rm QPO}$/$P_{\rm DNO}$=15.
From Warner et al.\ (2003).
\vspace{0.4cm}
}
}
\vspace{-0.1cm}
\end{figure}

\subsection{The CV Two-QPO Diagram}

The optical DNOs and QPOs in CVs behave in a fashion similar to what is
seen in the high and low frequency QPOs in LMXBs
(Psaltis et al.\ 1999), namely that 
$P_{\rm QPO}$$\sim$15$P_{\rm DNO}$ (Warner \&\ Woudt 2002; Mauche 2002a;
Fig.~\ref{qpodno}). 
The X-ray DNOs and QPOs in SS\,Cyg agree with this
relationship (Mauche 2002a), and the optical modulations in VW\,Hyi follow
such a relationship over a range of a factor of two as the star decreases in
brightness at the end of outburst (Woudt \&\ Warner 2002). Furthermore, the
CV `two-QPO diagram' (i.e., treating the DNOs as higher frequency QPOs) is
an extension of that in X-rays to frequencies three orders of magnitude
lower (Warner \&\ Woudt 2002; Warner et al.\ 2003; Fig.~\ref{qpodno}).

\section{X-ray emission from novae}
\label{novae}

\subsection{X-ray emission mechanisms}

In novae\footnote{For earlier X-ray reviews see, e.g.,  
\"Ogelman \&\ Orio (1995), 
Orio (1999), 
and Krautter (2002).}, after the thermo-nuclear runaway
in a H-burning shell at the bottom of the accreted layer on the WD,
a shock wave may appear and/or a strong radiation driven wind develops
(e.g., Webbink et al.\ 1987; 
Gallagher \&\ Starrfield 1978; 
Starrfield 1989; 
Shara 1989; 
Starrfield et al.\ 1990; 
Kato \&\ Hachisu 1994). 
Calculations show that not all the accreted material
is ejected, but a substantial fraction (10--90\%, depending on the system parameters)
can remain on the WD (e.g., Starrfield et al.\ 1972). 
Residual H burning of the remaining material in a shell on the WD
can thus take place which gives rise to radiation at a constant luminosity, 
$L_{\rm Edd}$ (referred to as the `constant bolometric
luminosity', CBL, phase). 
If not all of the accreted material is burned and blown away, 
the WD mass may increase towards the Chandrasekhar mass after a 
large number of outbursts in RNe, eventually leading to a type~Ia supernova event 
or to the formation of a neutron star by accretion induced collapse 
(e.g., Della-Valle \&\ Livio 1996; 
Orio \&\ Greiner 1999; 
but see Gonz\'alez-Riestra et al.\ 1998).

A short-lived X-ray emission phase ($\lesssim$1~day) is predicted when the 
energy of the burning shell first reaches the surface of the WD
(Starrfield et al.\ 1990).
The energy is then put into the envelope, which leads to a rapid expansion 
of the WD photosphere up to $\sim$100\,$R_{\odot}$; this leads to a drop 
in $T_{\rm eff}$.
As the WD photosphere subsequently shrinks back through the ejected material to the equilibrium radius of the WD
during the CBL phase, $T_{\rm eff}$ of the photosphere increases again
and the peak of the emitted spectrum shifts from visual to UV and finally to the
X-ray band (\"Ogelman et al.\ 1987, 1993; 
Shore et al.\ 1994;
MacDonald 1996; 
Balman et al.\ 1998, and references therein). 
At this stage the object is expected to radiate 
at $L_{\rm X}$$\sim$10$^{38}$\,erg\,s$^{-1}$ with
$kT_{\rm eff}$$\sim$20--85\,eV
(e.g., Starrfield 1979; Prialnik 1986), 
very much like a SSS (see Chapter 11). 
The duration of the CBL phase is expected to last a few years or less,
and is proportional to the left-over envelope
mass, which in turn is predicted to be inversely proportional to $M_{\rm WD}$
(e.g., Prialnik 1986; 
Starrfield 1989;
Kato \&\ Hachisu 1989, 1994; 
Kato 1997; 
see also Gonz\'alez-Riestra et al.\ 1998, and 
references therein).
Some novae do not show X-rays during the early stages of outburst.
This may be attributed to a large column density of the ejecta and argues in favour
of high masses for the expelled gas (e.g., Shore et al. 1996). 
Note that during the first part of the CBL phase the {\it observed} soft X-ray
flux may still increase, due to the clearing of the ejected material as it expands 
and its density decreases (Krautter et al.\ 1996). 
Once the H burning ceases, the WD photosphere is expected to cool at a constant
radius, and $L_{\rm X}$ drops gradually. 
Finally, when novae return to quiescence, accretion may resume again, and they
then appear as an ordinary CV. 
The total energy emitted during all stages is $\sim$10$^{44}$--10$^{46}$\,erg.

In addition to the soft component due to H-burning, 
shocks in the hot circumstellar material
can produce hard X-ray emission\footnote{Although we here refer to emission
typically $\gtrsim$0.1\,keV, we call it {\it hard} X-ray emission in order
to distinguish it from the soft component due to H-burning.}
(e.g., Brecher et al.\ 1977;
Willson et al.\ 1984; 
\"Ogelman et al.\ 1987;
O'Brien et al.\ 1994; 
Balman et al.\ 1998, and references therein). 
Even if there is no shock wave in the outburst, shocks may originate
in interacting winds, or they may be due to interaction between ejecta and 
pre-existing material. E.g., in RNe and
SBNe the donor is a giant 
star (often a Mira-type object in SBNe; see also Sect.~\ref{symbiotic}), 
which has lost a significant amount of material through a wind.
Note that CNe typically have main-sequence dwarf donors, which will not have 
significant stellar winds.
Alternatively, shocks may occur in the ejected shells during the different 
phases of the nova outburst.
The expected hard X-ray spectrum is thermal, with $kT_{\rm eff}$$\sim$0.2--15\,\,keV, 
depending on how much time has elapsed since the shock, and how efficient
the cooling is, and with
$L_{\rm X}$$\simeq$10$^{33}$--10$^{34}$\,erg\,s$^{-1}$
(e.g., Lloyd et al.\ 1992; 
O'Brien et al.\ 1994). 
Additionally a collection of emission lines are expected.
Compton scattering of $\gamma$-rays produced in
the decay of $^{22}$Na and $^{26}$Al to X-ray energies 
are expected to also give rise to hard X-ray emission (Starrfield et al.\ 1992; 
Livio et al.\ 1992; 
Pistinner et al.\ 1994). 
However, this may only become important at $kT$$\gtrsim$6\,keV
(Livio et al.\ 1992); moreover, they are probably not the main source
of hard X-rays (Starrfield et al.\ 1992). 

\subsection{Super-soft X-ray emission}
\label{supersoft}

The CN GQ\,Mus (Nova Mus 1983) was the first nova to show
supersoft emission similar to SSS (\"Ogelman et al.\ 1993). 
Subsequently, more novae were found to exhibit such a phase,
i.e., the CNe V1974\,Cyg (Nova Cyg 1992; Krautter et al.\ 1996; Balman et al.\ 1998), 
Nova LMC 1995 (Orio \&\ Greiner 1999),
V2487\,Oph (Nova Oph 1998; Hernanz \&\ Sala 2002), 
V382\,Vel (Nova Vel 1999; Orio et al.\ 2002), 
and V1494\,Aql (Nova Aql 1999 No.~2; e.g., Starrfield et al.\ 2001),
and the RN U\,Sco (Kahabka et al.\ 1999). 
The soft component has best-fit temperatures of
$kT$$\simeq$30--80\,eV with $L_{\rm X}$$\sim$10$^{37}$--10$^{38}$\,erg\,s$^{-1}$.
About 20\%\ of the observed novae have shown a supersoft phase
(although not all novae have been followed closely enough to exclude
such a phase). 
One of the reasons put forward why not all novae show a supersoft phase is,
that in most novae all the H might be depleted soon after the outburst,
therefore providing no time for a CBL phase (Orio et al.\ 2001a). 
Simple X-ray fits such as a black-body do not describe the soft component
well (see, e.g., Kahabka et al.\ 1999;
Balman \&\ Krautter 2001; Orio et al.\ 2002). 
This is because the spectral energy distribution of a 
hot WD differs considerably from a black-body
(e.g., MacDonald \&\ Vennes 1991; 
Jordan et al.\ 1994; 
Hartmann \&\ Heise 1997). 

The X-ray turn-off times for most CNe are generally $<$3--7 years
(e.g., Szkody \&\ Hoard 1994; 
Orio et al.\ 1996; 
Hernanz \&\ Sala 2002; 
see also Gonz\'alez-Riestra et al.\ 1998), 
suggesting that most of the CNe run out of nuclear fuel in the course of a few years
after outburst.
V1974\,Cyg (see Sect.~\ref{v1974cyg}) 
turned off only $\sim$18 months after outburst (Krautter et al.\ 1996).
The turn-off times are much shorter than the nuclear burning
times, and imply that the WD has ejected most of its
envelope during or soon after the outburst.
Note that GQ\,Mus turned off $\sim$9--10~years after outburst
(Shanley et al.\ 1995); 
at that time the temperature of the cooling remnant was $\sim$10\,eV.
This is somewhat longer than most novae, and it is argued
that either more mass was left on the WD than generally inferred, or that
the short $P_{\rm orb}$, 85.5\,min,
of GQ\,Mus and the small donor mass had enhanced 
the effects of irradiation of the donor (e.g., Diaz \&\ Steiner 1994). 
This in turn may have induced continued
accretion on the WD and prolonged the CBL phase
(\"Ogelman et al.\ 1993; 
Shanley et al.\ 1995). 

The RN CI\,Aql showed faint ($\sim$10$^{33}$\,erg\,s$^{-1}$)
and very soft ($kT$$\sim$40--50\,eV) X-ray emission, 14 and 16 months after
the outburst. However, this is not due to H-burning; that phase had already ceased before that time. 
The observed X-ray emission is suggested to be either due to ionization of the circumstellar
material or due to shocks within the wind and/or with the surrounding medium
(Greiner \&\ Di~Stefano 2002). 

In SBNe most of the accreted mass onto the WD is burned
quietly for decades, instead of being ejected away quickly as in most CNe.
Two SBNe showed supersoft X-ray spectra and luminosities 
(RR\,Tel: Jordan et al.\ 1994; 
M\"urset \&\ Nussbaumer 1994; 
SMC3/RX\,J0048.4$-$7332: Kahabka et al.\ 1994; 
Jordan et al.\ 1996). 

\subsection{Hard X-ray emission}

Hard X-ray emission attributed to shocked gas was first 
detected from the SBN RS\,Oph in its 1985 outburst.
The shocked gas is due to the expanding nova shell colliding
with the red giant wind
(e.g., Bode \&\ Kahn 1985). 
Most CNe and RNe emit hard X-rays in the first months of the
outburst with 
$L_{\rm X}$ peaking at $\sim$10$^{33}$--10$^{34}$\,erg\,s$^{-1}$ 
and bremsstrahlung temperatures in the range 0.5--20\,keV 
(e.g., Orio et al.\ 2001a, and references therein). 
The eclipsing, fast 
($t_3$$\simeq$3~days\footnote{$t_3$ is the time it takes for the nova to
decrease by 3 visual magnitudes.}) CN V838\,Her (Nova Her 1991)
was detected in hard X-rays 5 days after the outburst, with
a $kT$$\sim$10\,keV bremsstrahlung spectrum (Lloyd et al.\ 1992).
The hard emission generally lasts for at least $\simeq$2~years and even much longer under
special circumstances like pre-existing circumstellar
material, or a prolonged wind phase (e.g., Orio et al.\ 2001a).

SBNe show generally hard X-ray emission with 
$L_{\rm X}$$\lesssim$10$^{30}$--3$\times$10$^{33}$\,erg\,s$^{-1}$.
When plotted as time since outburst, there seems to be 
a general decay law for the X-ray flux in SBNe following outburst
(Allen 1981; 
Kwok \&\ Leahy 1984;
Hoard et al.\ 1996; 
M\"urset et al.\ 1997). 
This is consistent with the decay times of SBNe after the outburst of the order of decades.

\subsection{V1974\,Cyg and V382\,Vel}
\label{v1974cyg}

Two CNe, V1974\,Cyg and V382\,Vel, have been monitored quite frequently in
X-rays during the decline from their outburst.

\begin{figure*}[t]
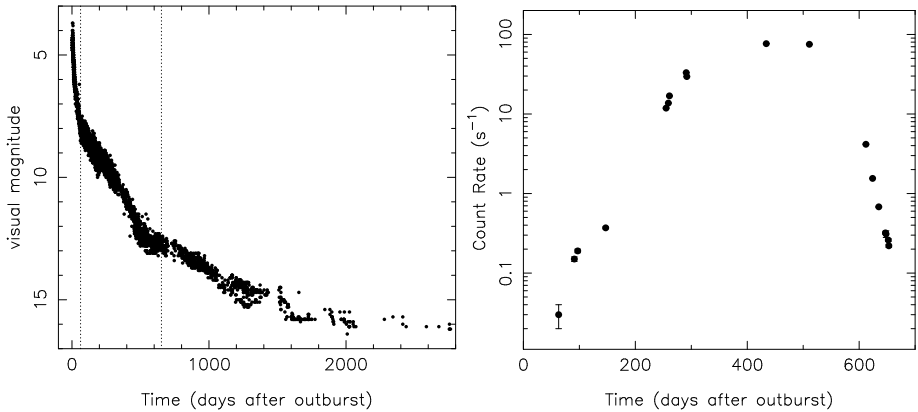

  \vbox{\hskip -\leftskip
  \hspace{0.2cm}
  \psfig{figure=knsw_fig8a.ps,width=6.0cm,angle=-90.0,%
  bbllx=135pt,bblly=50pt,bburx=589pt,bbury=550pt,clip=}
  \psfig{figure=knsw_fig8b.ps,width=6.0cm,angle=-90.0,%
  bbllx=135pt,bblly=43pt,bburx=589pt,bbury=550pt,clip=}}\par
  \caption{\label{krautter}
Optical light curve (({\it left}) based on VSOLJ, VSNET and AFOEV visual estimates and
X-ray light curve ({\it right}; from Krautter et al.\ 1996) based on {\it ROSAT}/PSPC data (0.1--2.4\,keV) 
of the CN V1974\,Cyg (Nova Cyg 1992). The start of the outburst is at Feb 19, 1992.
In the left panel we indicate with dotted lines the timespan of 
the right panel.
}
\end{figure*}

V1974\,Cyg was optically a bright nova 
(Fig.~\ref{krautter}).
It was the first nova to be observed at all wavelengths
from $\gamma$-rays to radio. 
It was a moderately fast nova ($t_3$$\sim$35~days) and it 
was followed from the early rise of the X-ray flux through the time of 
maximum to the turnoff (Krautter et al.\ 1996; Fig.~\ref{krautter}).
The initial observations showed a hard component with a peak around 1\,keV.
Subsequently, during the X-ray rise, a much softer component appeared that dominated the spectrum at maximum. This soft component decayed more rapidly 
(by a factor of $\sim$35) than the hard component.
During the early rise $N_{\rm H}$ decreased and then leveled off. 
The rise in the soft component flux
and the decrease in $N_{\rm H}$ is consistent with the clearance
of the nova ejecta. 
The X-ray spectra were best fitted by a WD atmosphere emission
model (soft component; $\lesssim$1\,keV) and a Raymond-Smith model of thermal plasma
(hard component; $\gtrsim$1\,keV), see Balman et al.\ (1998). 
The duration of the CBL is inferred to be $\gtrsim$511~days (Krautter et al.\ 1996).

The hard component spectrum evolved independently of the soft component.
The maximum of the hard X-ray emission was reached 
$\sim$150~days after the outburst with
$L_{\rm X}$$\sim$0.8--2$\times$10$^{34}$\,erg\,s$^{-1}$.
The time evolution of the hard X-ray flux and the fact that 
the plasma temperatures decreased 
from 10\,keV to 1\,keV suggest emission from shock-heated gas
(Balman et al.\ 1998).
This component arose from the interaction of the expanding nova wind
with density inhomogeneities, as is also indicated by {\it HST} images within the shell
(Paresce et al.\ 1995). 

UV observations showed that $\sim$500~days after the outburst, the ejecta
reached maximum ionization and then started to recombine
(Shore et al.\ 1996). 
This is coincident with the start of the X-ray turn off observed in X-rays. 
The decrease in the ionization fraction of the 
ejecta can be interpreted as a change in the photo-ionization rate from the 
central source and is due to a decrease in luminosity and $T_{\rm eff}$
after the cessation of nuclear burning. 
Two years after the X-ray turn-off a $kT_{\rm eff}$$\simeq$2\,eV 
(20\,000\,K) for the WD was inferred (Shore et al.\ 1997). 
It is possible that the initial X-ray decline in V1974\,Cyg was the result of a 
cooling WD in which most of the energy was radiated outside the X-ray band pass, 
rather than representing a return to final quiescence. This implies a longer cooling
time, hence a larger amount of mass remaining on the WD
(Shore et al.\ 1997).

V382\,Vel was a relatively fast nova ($t_3$$\sim$10\,d).
A short {\it RXTE} observation was performed $\sim$3 days after the peak
of the optical outburst, but no X-rays were seen.
Hard X-rays were detected $\sim$15--17~days after the peak of the optical 
outburst with $kT$$\sim$6--10\,keV (Orio et al.\ 2001b;
Mukai \&\ Ishida 2001). 
In the first two months after the outburst it cooled rapidly
to $kT$$\simeq$2.4\,keV (Mukai \&\ Ishida 2001).
L$_{\rm X}$ was $\sim$5$\times$10$^{34}$\,erg\,s$^{-1}$ for about 3--5 weeks.
As the initially strong intrinsic absorption 
($N_{\rm H}$$\sim$2$\times$10$^{23}$\,cm$^{-2}$; Orio et al.\ 2001b) of the ejected nebula
was thinning out, the equivalent $N_{\rm H}$ decreased to
$\sim$2$\times$10$^{22}$\,cm$^{-2}$ (Mukai \&\ Ishida 2001). 
Four months later the hard component had cooled
to a plasma temperature of $kT$$<$1\,keV and the absorption column was 
close to the interstellar value of $N_{\rm H}$$\sim$10$^{21}$\,cm$^{-2}$
(Orio et al.\ 2001b). 
The hard X-ray emission is interpreted as the result of a shock internal
to the nova ejecta.
The initial ejecta provide the absorbing column; a layer of later and faster moving
ejecta plough into the initial ejecta.
Note that in the first X-ray observations a weak Fe\,K line at $\simeq$6.6\,keV, with EW$\sim$130\,eV,
was seen. The weakness of the Fe\,K line is consistent with the shock model, provided
that the shocked plasma is not in ionization equilibrium (Mukai \&\ Ishida 2001). 

About six months after the outburst peak V382\,Vel appeared as 
a bright SSS (Orio et al.\ 2002). 
The supersoft X-rays were variable by a factor of $\sim$2 on a time scale of minutes, 
which are energy independent. The hard component 
did not show such variability.
Note that variability was also seen in soft X-rays 
from V1494\,Aql, about 10 months after the outburst: it showed a flare which lasted for 
$\sim$15\,min; 
additionally a periodicity near 42\,min was found
(Drake et al.\ 2003). 

Four months later, the continuum emission from V382\,Vel was gone; instead an emission line spectrum,
from highly ionised ions, in the supersoft range was inferred.
These emission lines presumably have their origin
from ionization within the ejected nebula.
Note that some of the spectra taken with relatively low
resolution may appear softer, if there is a strong superimposed nebular emission.
The WD then appears to be cooler than obtained from the spectral fits
(Orio et al.\ 2002).

V382\,Vel was followed for a further period of 8 months, 
starting 7 months after the outburst.
Within a period of less than 6 weeks the flux had dropped
by a factor $\sim$200. Thereafter this component continued to decline
more slowly, but still somewhat faster than the hard X-rays.
The spectra during this period show a wealth of emission lines, which mainly come from 
H and He-like transitions of Mg, Ne, O, N and C, as well as some Si and Na lines.
Most of the lines are broadened with FWHM$\sim$2000\,km\,s$^{-1}$, which is compatible
with the velocity of the expanding shell. The He-like triplets
of O\,{\sc VII} and N\,{\sc VI} constrain the plasma temperature to
$kT$$\simeq$39--43\,eV
(Burwitz et al.\ 2002b). 
One possibility is that during the late phase of the nova, when the shell has 
become transparent, the medium around the system is ionised by the still UV/EUV
bright WD. Alternatively, the line dominated spectrum could result from  the interaction
of the expanding shell with the circumstellar material, shocks within
the expanding shell, or from shocks due to collisions of a fast wind with the 
interstellar matter (Greiner \&\ Di Stefano 2002). 

\subsection{Quiescent, old novae}

Novae in quiescence are often referred to as `old' novae. 
Sample studies of old novae 
show X-ray emission with typically
$L_{\rm X}$$\simeq$10$^{30}$--10$^{33}$\,erg\,s$^{-1}$ and 
$kT$$\gtrsim$1\,keV,
when detected (e.g., Becker 1981, 1989; 
Becker \&\ Marshall 1981; 
C\'ordova \&\ Mason 1983; 
Orio et al.\ 1993, 2001;
Balman et al.\ 1995; 
\"Ogelman \&\ Orio 1995). 
This is comparable to that found for quiescent DNe
(see Sect.~\ref{overview}), although the optical quiescent 
novae are at least $\sim$10 times brighter.
Fast novae appear to be brighter than slow novae in quiescence 
(Becker \&\ Marshall 1981; Orio et al.\ 2001a).
Note that V2487\,Oph is the first nova to be detected in quiescence
both before and after the outburst, with similar fluxes (Hernanz \&\ Sala 2002).
The mass transfer rates derived from $L_{\rm X}$ are much lower
than those inferred from other wavelengths, similar to that found for 
DNe (see Sects.~\ref{quiescence}, \ref{outburst}).
In this case it might be that the BL radiation is emitted entirely in the EUV.
The ratio of $L_{\rm X}$ (0.2--2.4\,keV) to $L_{\rm opt}$ varies from 0.005 (DQ\,Her/Nova Her 1934) to 
4.8 (e.g., CP\,Pup/Nova Pup 1942); for the majority of the systems the value is
less than 0.01 (Orio et al.\ 2001a).

GK\,Per (Nova Per 1901) is the first nova for which a CN shell was detected in X-rays 
(Balman \&\ \"Ogelman 1999; Balman 2002). 
The X-ray nebula is asymmetric and composed of knots/clumps. 
The temperature and ionization structures do not vary much 
across the nebula. Its X-ray spectrum is thermal
with at least two temperature components, i.e., 
$kT$$\sim$0.2\,keV with $L_{\rm X}$$\sim$2$\times$10$^{31}$\,erg\,s$^{-1}$ 
and $kT$$>$30\,keV with $L_{\rm X}$$\sim$3$\times$10$^{31}$\,erg\,s$^{-1}$.
Distinct Ne\,{\sc IX} emission is seen.
The knots/clumps are thought to be the result of fragmentation and condensation
in the post-shock material. 
The existence of the X-ray shell in contrast to other CNe
is attributed to the high ambient density.

An outlier among the old novae is T\,Pyx. This RN showed 5 outbursts between 1890 and 1966.
Its outbursts are like those of slow novae, whereas all other RNe show fast nova outbursts.
Its luminosity in quiescence is higher than other quiescent novae, 
$L_{\rm X}$$\simeq$1--3$\times$10$^{36}$\,erg\,s$^{-1}$.
This has been attributed to steady nuclear burning on the WD 
(e.g., Webbink et al.\ 1987). 

\section{X-ray emission from symbiotic binaries}
\label{symbiotic}

SBs are divided into two subcategories based on their IR colours. S-type (`stellar') systems
have IR colours like those if isolated field red giants, whereas D-type 
(`dusty') systems have IR colours which are redder, indicative of dust.
D-type systems generally contain Mira variables with very high mass-loss rates.
The wind is ionised by the WD giving rise to the symbiotic nebula.

SBs show small `outbursts' where the optical increases by $\sim$1--2\,mag
on time scales of years.
The origin of these outbursts is still rather unclear. They may be related to
quasi-steady burning of matter on the WD, shell flashes, or unstable accretion.
Symbiotic Novae (SBNe) are a small subgroup of the SBs. 
They undergo large amplitude ($\sim$7\,mag increase) outbursts
with durations on the order of decades. These outbursts are thought 
to be due to thermonuclear runaway events on the surface of the WD.
SBNe are discussed in Sect.~\ref{novae}. 

Many of the SBs are detected in X-rays.
The most X-ray luminous are the D-types, whereas the S-types are typically 
two order of magnitudes fainter.
Also, SBNe are generally brighter in X-rays
compared to the other SBs. This is 
probably because the average luminosities of SBNe are higher than for the
other SBs.
SBs have been shown to emit either supersoft X-ray emission, and/or hard X-ray emission from 
an optically thin plasma with $kT$$\sim$0.25--1.3\,keV
with $L_{\rm X}$$\sim$10$^{30}$--10$^{33}$\,erg\,s$^{-1}$,
or even harder X-ray emission, as from an accreting neutron star
(see, e.g., M\"urset et al.\ 1997, and references therein).

The hard X-ray emission may be due to the colliding winds of the red giant and the WD 
(e.g., Willson et al.\ 1984; 
Kwok \&\ Leahy 1984),
jets, if they exist (e.g., Viotti et al.\ 1987; Leahy \&\ Volk 1995), jets colliding with
the interstellar material (e.g., Viotti et al.\ 1987; Ezuka et al.\ 1998), 
X-ray emission from the red giant (see, e.g., Ezuka et al.\ 1998), 
or accretion onto the WD (e.g., Leahy \&\ Taylor 1987; 
Jordan et al.\ 1994). 
The hard X-ray spectra can be interpreted as being composed of two components
(e.g., Leahy \&\ Volk 1995; 
Ezuka et al.\ 1998). However, Wheatley (2001) showed that the hard X-ray emission 
can be solely due to emission from the WD which is 
strongly absorbed by a partionally ionised wind from the red giant.

In SBs, Bondi-Hoyle capture of the red giant wind gives $\dot{M}$$\sim$10$^{-8}$\,yr$^{-1}$
onto the WD, which could produce nuclear burning in some 
systems (Sion \&\ Starrfield 1994). 
This might give rise to supersoft X-ray emission, similar to that seen 
in CNe and RNe (see Sect.~\ref{novae}) and other
SSS (see Chapter~11).
Depending on how much matter the WD receives and steadily burns, it may be able increase its 
mass. SBs have therefore been also put forward as candidates for Type~Ia supernovae
(e.g., Munari \&\ Renzini 1992). 
It has been noted, however, that either not enough mass can be accreted from the 
donor, or a mass outflow from the WD may inhibit such a phase 
(M\"urset \&\ Nussbaumer 1994; M\"urset et al.\ 1997).

Apart from two SBNe (RR\,Tel and SMC3, see Sect.~\ref{supersoft}), among the SBs only 
AG\,Dra has shown supersoft X-ray emission.
This system is also one of the brightest X-ray sources among the SBs
(e.g., Anderson et al.\ 1981). 
During quiescence X-ray spectra show a 
soft component with a temperature of $kT_{\rm bb}$$\simeq$14--15\,eV
(Greiner et al.\ 1997; see also Piro et al.\ 1985; Kenyon 1988). 
The X-ray luminosity remained constant during quiescence at a level
of $L_{\rm X}$$\simeq$10$^{37}$\,erg\,s$^{-1}$
(Greiner et al.\ 1997).

AG\,Dra is also the first SB to be observed during outburst in X-rays
(Viotti et al.\ 1995). 
During the minor ($\sim$1\,mag increase) and major ($\sim$2\,mag increase) outbursts 
AG\,Dra displays different behaviour.
Although during both outbursts the X-ray flux dropped, 
$kT$ decreased by a few eV
during the major outbursts,
while it increased by a few eV during the minor outbursts.
The behaviour during the major outbursts is explained by expansion 
(factor 2--6) and cooling 
of the WD atmosphere, which also explains the anti-correlation between optical/UV and X-ray fluxes
(Greiner et al.\ 1997; Gonz\'alez-Riestra et al.\ 1999).
The cooling could be the result of an increased $\dot{M}$ onto the compact object
causing it to expand slowly (Greiner et al.\ 1997).
The minimum X-ray flux during `hot' outbursts might be attributed to the increased
absorbing layer between the X-ray source and observer
(Friedjung 1988).

\section{Concluding remarks}

During the last decade there has been a staggering flow of new information, not
only in X-rays, but also from other wavelengths.  Nowadays, substantial
progress in understanding the behaviour of CVs is generally made through
multi-wavelength efforts. Unfortunately, the space available in this Chapter
has not allowed us to fully discuss observations across the whole
electro-magnetic spectrum. Nevertheless, we have shown that our somewhat
`restricted' view at EUV and X-ray wavelengths has revealed a
great variety of rich phenomena taking place in the environment of an accreting
white dwarf.

Substantial improvements in our understanding of CVs at high energies are
already coming from the instruments on board {\it Chandra} and {\it XMM-Newton}
and further advances are inevitable over the next few years. These instruments
provide the necessary spectral resolution to perform sensible temperature and
velocity diagnostics from individually resolved lines and line ratios;
something we have previously been used to at optical wavelengths.  These
exquisite instruments and those which are yet to come will provide us with even
more information, and will leave us enough work for many years to come.

\subsection*{Acknowledgements}
We acknowledge discussions with K.~Beuermann, J.-P.~Lasota, C.~Mauche, 
M.~Orio, S.~Starrfield and P.~Wheatley.
AS is supported by the German 
Bundesministerium f\"ur Bildung und Forschung through 
the Deutsches Zentrum f\"ur Luft- und Raumfahrt e.V.\ (DLR) 
under grant number 50\,OR\,9706\,8. BW is supported 
by research funds from the University of Cape Town.

\end{document}